\documentclass[
    prd, twocolumn, superscriptaddress, nofootinbib, amsmath, amssymb,
    aps, floatfix, preprintnumbers
]{revtex4-2}
\usepackage[utf8]{inputenc}
\usepackage{setspace}
\usepackage[dvips]{graphicx}
\usepackage{booktabs}
\usepackage[dvipsnames]{xcolor}
\usepackage{tikz}
\usepackage{pifont}
\usepackage{ulem}
\definecolor{CiteBlue}{RGB}{45,52,151}
\usepackage[
    colorlinks=true,
    linkcolor=CiteBlue,
    urlcolor=CiteBlue,
    citecolor=CiteBlue
]{hyperref}

\usepackage[version=4]{mhchem}
\usepackage{aas_macros}

\usepackage[capitalise]{cleveref}
\usepackage{siunitx}
\DeclareSIUnit{\year}{yr}
\usepackage{physics}
\usepackage{slashed}
\usepackage{soul}
\newcommand{\nn}{\nonumber}

\usepackage{bm}

\newcommand{\du}{\mathrm{d}}
\renewcommand{\dd}{\,\du}

\renewcommand{\Im}{\operatorname{Im}}

\newcommand{\el}{\mathrm{e}}

\newcommand{\CF}{\hat{{\mathcal{F}}}}

\newcommand{\newrow}[1]{\nn \, {#1} \\[5pt]}

\newcommand{\SIff}{F_0}
\newcommand{\SDff}{\vb{F}}
\newcommand{\generalV}{V(\vb{q} ,\vb{v}_\perp)}

\definecolor{darkgreen}{rgb}{0,0.45,0}

\newsavebox{\twosubbox}

\begin{document}
\setstcolor{magenta}
\title{Determining (All) Dark Matter--Electron Scattering Rates From Material Properties}

\author{Yonit Hochberg}
\affiliation{Racah Institute of Physics, Hebrew University of Jerusalem, Jerusalem 91904, Israel}
\affiliation{Laboratory for Elementary Particle Physics,
 Cornell University, Ithaca, NY 14853, USA}

\author{Majed Khalaf}
\affiliation{Racah Institute of Physics, Hebrew University of Jerusalem, Jerusalem 91904, Israel}
\affiliation{Laboratory for Elementary Particle Physics,
 Cornell University, Ithaca, NY 14853, USA}

 \author{Alessandro Lenoci}
\affiliation{Racah Institute of Physics, Hebrew University of Jerusalem, Jerusalem 91904, Israel}
\affiliation{Laboratory for Elementary Particle Physics,
 Cornell University, Ithaca, NY 14853, USA}

\author{Rotem Ovadia}
\affiliation{Racah Institute of Physics, Hebrew University of Jerusalem, Jerusalem 91904, Israel}
\affiliation{Laboratory for Elementary Particle Physics,
 Cornell University, Ithaca, NY 14853, USA}

\date\today

\begin{abstract}\ignorespaces{}
    We show that the scattering rate for any dark matter (DM) interaction with electrons in 
    any target is proportional to several measurable material properties, encapsulated by a single master formula.
    This generalizes the dielectric function formalism---developed for DM interactions that couple to electron density---to any interaction, incorporating both spin-dependent and spin-independent interactions simultaneously.     
    This formalism links the full many-body response of a target system to the DM probe in a clear and simple form, providing a reliable event rate prediction from measurable material quantities.
    We demonstrate the utility of our  formalism by placing new limits from existing data on a class of spin-dependent light DM interactions,  as their rates---contrary to common lore---are determined entirely by the dielectric function.
    We further highlight a promising avenue for the detection of sub-MeV DM using the rare earth metal Praseodymium, which exhibits a spin-dependent anisotropic response down to the ${\rm meV}$ scale. 
    Our results lay the groundwork for a rapid systematic investigation of novel electron scattering targets going beyond the classic spin-independent searches, enhancing the prospects for DM detection.
\end{abstract}

\maketitle

%%%%%%%%%%%%%%%%%%%%%%%%%%%%%%%%%%%%%%%%%
\section{Introduction} \label{sec:intro}
%%%%%%%%%%%%%%%%%%%%%%%%%%%%%%%%%%%%%%%%%
%
The past decade has seen substantial progress in devising strategies for laboratory detection of dark matter~(DM). 
In particular, DM interactions with electrons offer an incredible opportunity to detect light DM of sub-GeV and even sub-MeV mass, with a growing variety of proposed target systems and detector designs~\cite{Essig:2011nj,Graham:2012su,Essig:2015cda,Hochberg:2015pha,Hochberg:2015fth,Hochberg:2019cyy,Griffin:2019mvc,Hochberg:2021yud,Derenzo:2016fse,Hochberg:2016ntt,Hochberg:2017wce,Cavoto:2017otc,Kurinsky:2019pgb,Canonica:2020omq,Kim:2024xea,Simchony:2024kcn,Blanco:2022cel,Essig:2022dfa,Das:2022srn,Das:2024jdz,Hochberg:2025dom,Chou:2023hcc,Du:2022dxf,Marocco:2025eqw,Schwemmbauer:2025evp,Griffin:2024cew}. 
Recent measurements~\cite{SuperCDMS:2020aus,SENSEI:2023zdf, Baudis:2025zyn, TESSERACT:2025tfw, DAMIC-M:2025luv} are probing new territory in light DM parameter space, with experimental energy thresholds reaching as low as $\sim 113 \, {\rm meV}$~\cite{Baudis:2025zyn}, corresponding to DM masses as low as a few tens of keV. 
With the ongoing incorporation of mature quantum sensing technologies, energy thresholds are expected to continue to rapidly decrease towards the meV scale in the upcoming decade.
This in turn should allow the DM direct detection program to capitalize on a variety of target materials that exhibit emergent multi-body excitations at low energies.

Key to the success of this program is the ability to directly link physical response properties of a material to its prospects to detect DM.  
Refs.~\cite{Hochberg:2021pkt,Knapen:2021run,Boyd:2022tcn} used linear response theory to show that for  DM interactions with the electron density, such as scalar- or vector-mediated DM-electron interactions, the entire multi-body detector response to the DM probe is encapsulated by the dielectric tensor of the material. 
This allowed for the rapid calculation of the 
spin-independent
DM scattering rate off a given target from a physics-observable perspective, enabling one to go beyond a single-particle excitation description, avoiding approximations inherent to analytical modeling, and facilitating high-throughput searches for optimal materials~\cite{Griffin:2025wew}.

Going beyond spin-independent interactions has been an active area of research~\cite{Trickle:2019ovy,Trickle:2020oki,Catena:2024rym, Marocco:2025eqw, Berlin:2025uka, Catena:2025sxz}.
In this work, we expand the dielectric formalism to encompass {\it all} non-relativistic DM-electron interactions, showing that the detector response is governed by 
several experimentally measurable material response functions: (i) the dielectric tensor, (ii) the electronic spin susceptibility (closely related to the magnetic susceptibility), and (iii) the charge-spin response.
Our result is provided in the form of a compact and simple master formula, Eq.~\eqref{eq:master_eq}, 
that explicitly lays out the relation between material responses and the event rate from an ambient non-relativistic DM distribution.
Our master formula provides
non-trivial insights that have been overlooked so far:
\begin{itemize}
    \item Importantly, there are \emph{spin-dependent} interactions between DM and electrons in which the \emph{dielectric tensor} alone controls the scattering rate: these are interactions which depend on the DM spin but not on the electron spin.
    Among the three material response functions, dielectric data is by far the most readily available, highlighting the utility of this result. 
    We explicitly demonstrate this by using experimental data from the QROCODILE~\cite{Baudis:2025zyn} and DAMIC-M~\cite{DAMIC-M:2025luv} collaborations to recast their spin-independent bounds and place the first direct detection constraints on {\it e.g.} electric dipole and anapole DM~\cite{Ho:2012bg} scattering off electrons. 
    Similarly, we establish the future reach of existing proposals geared at spin-independent interactions into this complementary DM parameter space. 

    \item Existing literature on spin-dependent interactions, where the electron spin participates in the interaction with DM, has primarily focused on detection via magnon excitations in ferromagnetic and antiferromagnetic materials~\cite{Trickle:2019ovy,Trickle:2020oki,Marocco:2025eqw,Berlin:2025uka}, thanks to the simple analytic modeling of their anisotropic response. 
    Our master formula places all materials and excitations on equal footing, thereby elucidating that other materials exhibiting a variety of multi-body excitations, such as paramagnets, can provide exceptional sensitivity.
    We illustrate this for a commercially available Pr crystal, which exhibits a strong anisotropic response and a projected reach into DM parameter space that surpasses existing proposals by several orders of magnitude.
    Our results motivate extending the search for optimal detector materials beyond those exhibiting magnonic excitations, to include {\it e.g.} paramagnets. 
\end{itemize}

This paper is organized as follows. Section~\ref{sec:master} presents our master formula for DM-electron scattering rates. In Section~\ref{sec:demo} we demonstrate the implications via several examples. Section~\ref{ssec:spin-dependent from dielectric} shows how to use dielectric data to constrain spin-dependent DM-electron interactions, and places new limits on the relevant parameter space from existing experimental data. Section~\ref{ssec:spin} focuses on the electronic spin response, highlighting the reach of Pr into the relevant DM parameter space. We conclude in Section~\ref{sec:sum}. 
A set of appendices contains further details on the formalism developed in this work, as well as auxiliary calculations useful to reproduce our results. Appendix~\ref{app:operator NR limit} shows the mapping between the set of four-Fermion interactions and the non-relativistic operator basis. 
A brief review of linear response theory, the derivation of our master formula and the explicit mapping to the dielectric function formalism are detailed in Appendices~\ref{app:linear response theory},~\ref{app:master eq derivation} and \ref{app:mapping to dielectric}, respectively. 
Appendix~\ref{app:Praseodymium} provides additional details regarding the Pr spin response.
Finally, phase space integrals used in our rate calculations are given in Appendix~\ref{app:auxiliary calculations}.

%%%%%%%%%%%%%%%%%%%%%%%%%%%%%%%%%%%%%%%%%%%%%
\section{Master Formula} \label{sec:master}
%%%%%%%%%%%%%%%%%%%%%%%%%%%%%%%%%%%%%%%%%%%%%
%
The full electronic detector response to a weak DM probe of any type is completely accounted for by a few physical quantities. 
Each one is either measurable or numerically calculable and characterizes the response of the electron charge and spin densities to weak~external~forces:
\begin{itemize}
    \item 
    The charge density response $\chi_{00}$---related to the dielectric tensor of the material---has been shown~\cite{Hochberg:2021pkt, Knapen:2021run, Boyd:2022tcn} to account for the entire detector response in the case of spin-independent interactions, and can be reliably measured through electron scattering or photon absorption measurements.
    Explicitly, ${\rm Im}(-\chi_{00}) = (\vb{q}^2 / e^2) {\rm Im}\left(-\epsilon_L^{-1}\right)$, where $\vb{q}$ is the transferred momentum, $\epsilon_L$ is the longitudinal projection of the dielectric tensor along $\vu{q}$, and $e$ is the electromagnetic coupling.
    It can be measured by various probes, {\it e.g.} using infrared spectroscopy, X-ray scattering and electron energy-loss spectroscopy (EELS).
    \item 
    The spin-density response $\chi_{ij}$, also called the spin-susceptibility, discussed in several recent works~\cite{Catena:2024rym, Marocco:2025eqw, Berlin:2025uka, Catena:2025sxz}, is a tensor describing how the electronic spin density of the material changes in response to the forces acting on it.
    The spin-susceptibility is the main contribution to the magnetic susceptibility tensor $\chi^m_{ij}$, with the relation between them given by $\chi^m_{ij} \sim \mu_B^{2} \chi_{ij} + O(\mu_N / \mu_B)$, where $\mu_B\, (\mu_N)$ is the Bohr (nuclear) magneton and the sub-leading contributions come from nuclear spin responses.
    Recent direct detection literature has primarily focused on ferromagnets or anti-ferromagnets, where the spin-susceptibility was approximated by a magnon model~\cite{Trickle:2019ovy,Marocco:2025eqw,Berlin:2025uka}.
    The spin-susceptibility can be measured in neutron scattering experiments~\cite{Boothroyd:2020,Berlin:2025uka} with large databases available for high throughput exploration such as the ISIS INS database~\cite{ISIS_DataGateway}.
    \item 
    The spin-charge and charge-spin responses, denoted by  $\chi_{i0}$ and $\chi_{0i}$ respectively, are related to changes in the electron spin density as a result of forces acting on the electron charge density, and vice versa. 
    These responses emerge from couplings between the electron charge and spin degrees of freedom, {\it e.g.} through a spin-orbit interaction (Rashba effect~\cite{Bychkov:1984}). 
    They are at the heart of several contemporary areas of research in condensed matter physics, including the spin-hall effect and topological insulators~\cite{Manchon:2015}, where these responses are large. 
    In the context of DM detection, they have been shown to induce meV-scale band gaps in materials~\cite{Inzani:2020szg}.   
    $\chi_{i0}$ and $\chi_{0i}$ can be measured using a variety of probes, and are related to the spin-hall conductivity which has been the topic of several recent high-throughput studies~\cite{Meinert:2020, Zhang:2021}.
    Notably, the spin-charge response is the key to the maturing field of spintronics, suggesting an exciting opportunity for the incorporation of this technology into the direct detection program. 
\end{itemize}

We now demonstrate 
the relevance of these physical material properties to DM scattering. 
Consider a non-relativistic spin $1/2$ DM particle $\chi$ of mass $m_\chi$, which elastically scatters with electrons in a target at rest, depositing energy $\omega$ and momentum $\vb{q}$ in the process.
The small velocity is motivated by models for the DM in the galactic halo~\cite{Lewin:1995rx} for which the typical DM velocities are of order $\sim 10^{-3}$. 
Kinematics dictates that the transferred energy is $\omega_{\vb{q}} \equiv \vb{q} \cdot \vb{v} - q^2 / 2m_\chi$, where $q\equiv |\vb{q}|$, and $\bf v$ is the DM velocity.
We describe the DM-electron interaction using a non-relativistic effective theory~\cite{Fan:2010gt, Fitzpatrick:2012ix, Gresham:2014vja} where all operators are constructed from the Galilean invariants $\vb{q}$, $\vb{v}_\perp$, $\vb{S}_\chi$ and $\vb{S}_e$. Here, $\vb{v}_\perp$ is the component of $\bf v$ that is perpendicular to $\vb{q}$, $\vb{S}_\chi$ is the DM spin operator, and $\vb{S}_e$ is the electron spin density.

The most general DM-electron interaction Hamiltonian density in momentum space is given by
\begin{equation} \label{eq:Hint}
    {\cal H}_{\rm int}(\vb{q}) \equiv \generalV \,\Big( \SDff\cdot \vb{S}_e(\vb{q}) + \SIff\,n_e\pqty{\vb{q}} \Big) \, ,
\end{equation}
which is at most linear in the DM and electron spins.
This is generically true for spin-1/2 particles.
The factor $V(\vb{q},\vb{v}_\perp)$ is a spin-independent function that sets the energy scale of the interaction. 
More commonly, the less general parameterization $V(q)$ is used as it is appropriate for most well-motivated beyond Standard Model scenarios such as interactions mediated by bosonic mediators.
For example, if the interaction is mediated by a scalar of mass $m_\phi$ one finds $V(q) \sim  g_{e\chi}(q^2 + m_\phi^2)^{-1}$ where $g_{e\chi}$ is the effective DM-electron coupling.
$\vb{S}_{e}\pqty{\vb{q}}$ and $n_{e}(\vb{q})$ are the Fourier transform of the electron target spin and number density operators. 
$\SDff$ and $\SIff$ are operators  that generally depend on  $\vb{q}$, $\vb{v}_\perp$, and $\vb{S}_\chi$, with the explicit dependence suppressed for brevity.  
They are defined as
\begin{eqnarray}\label{eq:form factors}
    \SDff  &= &\frac{\partial {\cal O}}{\partial \vb{S}_e} \, , \qquad 
    \SIff  = {\cal O} - \SDff \cdot \vb{S}_e \, ,
\end{eqnarray}
where ${\cal O} = {\cal H}_{\rm int} / V(\vb{q}, \vb{v}_\perp)$ is dimensionless and accounts for any additional structure of the interaction such as spin and velocity dependence. 
One can span ${\cal O}$ using the well-established basis of fifteen non-relativistic operators~\cite{Fan:2010gt, Fitzpatrick:2012ix, Gresham:2014vja} presented in Table~\ref{tab:NR_operators}.
A mapping between relativistic dimension six, seven, and eight four-Fermi operators and the non-relativistic basis in Table~\ref{tab:NR_operators} can be found in Refs.~\cite{Fan:2010gt, Fitzpatrick:2010br} and is also provided in Table~\ref{table:DM-electron relativistic operators} in Appendix~\ref{app:operator NR limit}. Note that while both terms in Eq.~\eqref{eq:Hint} can depend on the DM spin, only the first term depends on the \emph{electron} spin.

%%%%%%%%%%%%%%%%%%%%%%%%%%%%%%%%%%%%%%%%%%%%%%%%%%%%%%%%%%%%%%%%%%%%%%%%%%%
\begin{table*}[ht]
    \centering
    \scriptsize 
    \begin{tabular}{|c@{\hskip 10pt} c@{\hskip 10pt}|c@{\hskip 10pt} c@{\hskip 10pt}|c@{\hskip 10pt} c@{\hskip 10pt}|c@{\hskip 10pt} c@{\hskip 10pt}|}
    \hline\hline
    {\bf Name} & {\bf Operator} & {\bf Name} & {\bf Operator} & {\bf Name} & {\bf Operator} & {\bf Name} & {\bf Operator} \\
    \hline
    $\mathcal{O}_1$ & $\vb{1}$ &
    $\mathcal{O}_5$ & $i\vb{S}_\chi \cdot \pqty{\dfrac{\vb{q}}{m_\chi} \times \vb{v}_{\perp}}$ &
    $\mathcal{O}_9$ & $i\vb{S}_\chi \cdot \pqty{\vb{S}_e \times \dfrac{\vb{q}}{m_e}}$ &
    $\mathcal{O}_{13}$ & $i\pqty{ \vb{S}_\chi \cdot  \vb{v}_{\perp}}\pqty{  \vb{S}_e \cdot \dfrac{\vb{q}}{m_e}}$ \\

    $\mathcal{O}_2$ & $v_{ \perp}^2$ &
    $\mathcal{O}_6$ & $\pqty{\vb{S}_\chi \cdot \dfrac{\vb{q}}{m_\chi}} \pqty{\vb{S}_e \cdot \dfrac{\vb{q}}{m_e}}$ &
    $\mathcal{O}_{10}$ & $i\vb{S}_e \cdot \dfrac{\vb{q}}{m_e}$ &
    $\mathcal{O}_{14}$ & $i\pqty{ \vb{S}_\chi \cdot \dfrac{\vb{q}}{m_\chi} }\pqty{  \vb{S}_e \cdot \vb{v}_{\perp}}$ \\

    $\mathcal{O}_3$ & $i \vb{S}_e \cdot \pqty{\dfrac{\vb{q}}{m_e} \times \vb{v}_{\perp}}$ &
    $\mathcal{O}_7$ & $\vb{S}_e \cdot \vb{v}_{\perp}$ &
    $\mathcal{O}_{11}$ & $i\vb{S}_\chi \cdot \dfrac{\vb{q}}{m_\chi}$ &
    $\mathcal{O}_{15}$ & $-\pqty{ \vb{S}_\chi \cdot \dfrac{\vb{q}}{m_\chi} }\pqty{  \vb{S}_e  \times \vb{v}_{\perp}\cdot \dfrac{\vb{q}}{m_e}}$ \\

    $\mathcal{O}_4$ & $\vb{S}_\chi \cdot \vb{S}_e$ &
    $\mathcal{O}_8$ & $\vb{S}_\chi \cdot \vb{v}_{\perp}$ &
    $\mathcal{O}_{12}$ & $\vb{S}_\chi \cdot \pqty{\vb{S}_e \times \vb{v}_{\perp}}$ &
    & \\ 
    \hline\hline 
    \end{tabular}
    \caption{{\bf Operators.} Basis of non-relativistic operators suppressed at most by the DM velocity squared, {\it i.e.}\ ${\bf q}/m_\chi$ or $\bf v_\perp$ squared, and linear in DM and electron spins. Operators involving higher powers of the DM velocity can be expressed as polynomials in this basis.
    Note that in this table,  the DM spin $\vb{S}_\chi$ should be treated as a quantum mechanical operator whereas $\vb{q}, \vb{v}_\perp$ as c-numbers (eigenvalues of their respective operators).
    }
    \label{tab:NR_operators}
\end{table*}
%%%%%%%%%%%%%%%%%%%%%%%%%%%%%%%%%%%%%%%%%%%%%%%%%%%%%%%%%%%%%%%%%%%%%%%%%%%%%%

The DM scattering rate per unit detector mass per unit exposure time
is given by 
\begin{align}\label{eq:R formula}
    R = \frac{1}{\rho_{\rm T}}\frac{\rho_\chi}{m_{\chi}} \frac{\pi \bar {\sigma}_e}{\mu_{e\chi}^2 \ev{\abs{V(\vb{q}^{\rm ref}, \vb{v}^{\rm ref}_\perp)}^2}_{\Omega}} \int \dd^3 {\bf v}\, f({\bf v}) \Gamma_{\cal{O}}({\bf v})  \, ,
\end{align}
where
\begin{align}\label{eq:Gamma v}
    \Gamma_{\cal{O}}({\bf v}) =\int\frac{\dd\omega \dd^3 {\bf q} }{(2\pi)^4} \frac{\dd\Gamma_{\cal{O}}}{\dd^3\vb{q}\dd\omega}
\end{align}
is the velocity-dependent scattering rate of a single DM particle with the target electrons.
In the prefactor, $\rho_{\rm T}$ is the target mass density, $\rho_\chi\simeq 0.4 \, {\rm GeV} / {\rm cm}^{3}$ is the DM density, $\mu_{e \chi}$ is the DM-electron reduced mass and $\bar{\sigma}_\el \equiv (\mu_{e \chi}^2 / \pi)  \ev{\abs{V(\vb{q}^{\rm ref}, \vb{v}^{\rm ref}_\perp)}^2}_{\Omega}$ is the fiducial reference cross-section
where $q^{\rm ref} = m_\chi v^{\rm ref} = \alpha m_e$, with $q^{\rm ref}$ denoting the modulus of the reference momentum $\vb{q}^{\rm ref}$, $\vb{v}^{\rm ref}$ its corresponding reference velocity, and $\ev{\cdot}_\Omega$ denoting averaging over all directions of $\vu{q}^{\rm ref}, \vu{v}^{\rm ref}$, assuming a uniform distribution. In Eq.~\eqref{eq:R formula} $f(\vb{v})$ is the DM velocity distribution, with the integral representing the velocity-averaged scattering rate, and in Eq.~\eqref{eq:Gamma v} $\frac{\dd\Gamma_{\cal{O}}}{\dd^3\vb{q}\dd\omega}$ is the differential DM scattering rate per energy-momentum deposit of~($\omega, \vb{q})$.

Using linear response theory 
one finds that as long as the DM-electron coupling is weak, one can express the differential rate in terms of the aforementioned measurable material responses.
The relation is given by our master formula as follows:
\begin{align} \label{eq:master_eq} 
    \frac{\dd\Gamma_{\cal{O}}}{\dd^3\vb{q}\dd\omega} &= \abs{\generalV}^2 (1+f_{\rm BE}(\omega))\, (2\pi) \, \delta(\omega-\omega_{\vb{q}})\nonumber\\
    \times&
    \Bigg[{\rm{Im}}\left(-\chi_{00} (\omega, {\bf q})\right)\,{\rm{Tr}}\Big\{ \SIff  \SIff^\dagger\Big\}\nonumber\\
    &+{\rm{Im}}\left(-\chi^+_{ij} (\omega, {\bf q})\right)\,{\rm{Re}}\left({\rm{Tr}}\Big\{ F_i   F_j^\dagger\Big\}\right)\nonumber\\
    &+{\rm{Re}}\left(-\chi^{-}_{ij} (\omega, {\bf q})\right)\,{\rm{Im}}\left({\rm{Tr}}\Big\{ F_i  F_j^\dagger\Big\}\right)\nonumber\\
    &+2\,{\rm{Im}}\left(-\chi^+_{0i} (\omega, {\bf q})\right)\,{\rm{Re}}\left({\rm{Tr}}\Big\{ \SIff  F_i^\dagger\Big\}\right)\nonumber\\
    &+2\,{\rm{Re}}\left(-\chi^{-}_{0i} (\omega, {\bf q})\right)\,{\rm{Im}}\left({\rm{Tr}}\Big\{ \SIff F_i^\dagger\Big\}\right)\Bigg]\,.
\end{align}
The first line of Eq.~\eqref{eq:master_eq} contains the stimulated emission factor $(1+f_{\rm BE})$ and the Dirac delta function enforcing the kinematics of the interaction, whereas the other lines are expressed in terms of the material charge and spin responses $\chi^{\pm}_{\mu\nu}\equiv (\chi_{\mu\nu}\pm\chi_{\nu\mu})/2$ where $0 \leq \mu,\nu \leq 3$. 
The trace denotes 
averaging over initial and summing over final DM spins, where the DM spin distribution is assumed to be uniform.
The stimulated emission factor typically evaluates to unity in the context of direct detection and will be dropped henceforth.  
Eq.~\eqref{eq:master_eq} is obtained using linear response theory and the fluctuation dissipation theorem. 
A comprehensive derivation is presented in Appendices~\ref{app:linear response theory} and \ref{app:master eq derivation}. (In this context, see also Refs.~\cite{Trickle:2020oki,Catena:2024rym,Catena:2025sxz, Chen:2025cvl}.)

The master formula Eq.~\eqref{eq:master_eq} explicitly shows the factorization of the rate into a material dependent part~$\sim\chi$ and DM model dependent part~$\sim\abs{V}^2 F F^\dagger$ for any type of interaction, providing a comprehensive recipe for evaluating any DM scattering rate.
Through the rapid evaluation of the full response of materials, one can easily compare vastly different DM detectors on equal footing. 
In many phenomenologically relevant cases, a subset or even a single material response may be sufficient to determine the DM rate.
For example, for any interactions where $\vb{F}=0$, namely the ones that do not depend on the electron spin, the interaction rate is entirely determined by ${\rm Im}(-\chi_{00})$,  {\it i.e.} by the dielectric function.
We emphasize that this includes a class of {\it spin-dependent} interactions that depend on the DM spin but not on the electron spin density. 
As we demonstrate in Section~\ref{ssec:spin-dependent from dielectric}, this enables the recasting of spin-independent direct detection results~\cite{DAMIC-M:2025luv, Baudis:2025zyn} to the complementary parameter space of several spin-dependent interactions.
For DM interactions (linearly) proportional to the electron spin density $\vb{S}_e$, one finds $F_0 = 0$, thereby the spin susceptibility $\chi_{ij}$ solely determines the interaction rate.
The terms ${\rm Re}(-\chi^-_{ij})$, ${\rm{Re}}\left(-\chi^-_{0i}\right)$ and ${\rm{Im}}\left(-\chi^+_{0i}\right)$  contribute to the DM scattering rate only when combinations of different operators ${\cal O} = \sum_i c_i {\cal O}_i$ are considered (for more details, see the end of Appendix~\ref{app:master eq derivation}).

%%%%%%%%%%%%%%%%%%%%%%%%%%%%%%%%%%%%%%%%
\section{Examples}\label{sec:demo}
%%%%%%%%%%%%%%%%%%%%%%%%%%%%%%%%%%%%%%%%
%
Having presented the master formula Eq.~\eqref{eq:master_eq}, we now highlight several of its features. 
Eq.~\eqref{eq:master_eq} provides a straightforward way to calculate the scattering rate in any material for any given DM-electron interaction---be it spin-dependent, spin-independent, or any combination of both types of interactions. 
In particular, it allows calculating the rate of some {\it spin-dependent} interactions using the {\it dielectric function} alone. 
The latter is a powerful result, since the dielectric function is the most accessible material response out of the three, where simple analytical models are readily available, along with abundant experimental measurements through electric probes.
This further allows one to recast the entire existing body of direct detection literature for spin-independent interactions based on the dielectric function to additional DM-electron spin-dependent interactions. The clear and direct relationship between material response and the DM interaction rate further allows one to identify new classes of materials of interest exhibiting strong responses, that have thus far been overlooked.

In what follows, we demonstrate these properties.
In all examples, we take $V(q) = g_{\chi e}(q^2 + m_\phi^2)^{-1}$, which could arise from a scalar or vector mediator of mass $m_\phi$ generating a coupling $g_{\chi e}$ between the DM and electrons.
The DM velocity distribution is assumed to follow the Standard Halo Model with $v_0 = 220\, {\rm km}/{\rm s}$, $v_{\oplus} = 232 \, {\rm km}/{\rm s}$,  $v_{\rm esc} = 540 \, {\rm km}/{\rm s}$~\cite{Lewin:1995rx,Stanic:2025yze}, and the DM spin distribution taken to be uniform.
Our statistical analysis assumes the DM scattering events to be distributed according to a Poisson distribution, with all new limits and projections made at the 95\% C.L.~\cite{Feldman:1997qc}.

%%%%%%%%%%%%%%%%%%%%%%%%%%%%%%%%%%%%%%%%%%%%%%%%%%%%%%%%%%%%%%%%%%%%%%%%%%%%%%%%%
\subsection{Spin-Dependent Results from the \\ Dielectric Function }\label{ssec:spin-dependent from dielectric}
%%%%%%%%%%%%%%%%%%%%%%%%%%%%%%%%%%%%%%%%%%%%%%%%%%%%%%%%%%%%%%%%%%%%%%%%%%%%%%%%%
%

%%%%%%%%%%%%%%%%%%%%%%%%%%%%%%%%%%%%%%%%%%%%%%%%%%%%%%%
\begin{figure*}[ht!]
    \centering
    \includegraphics[width=0.98\linewidth]{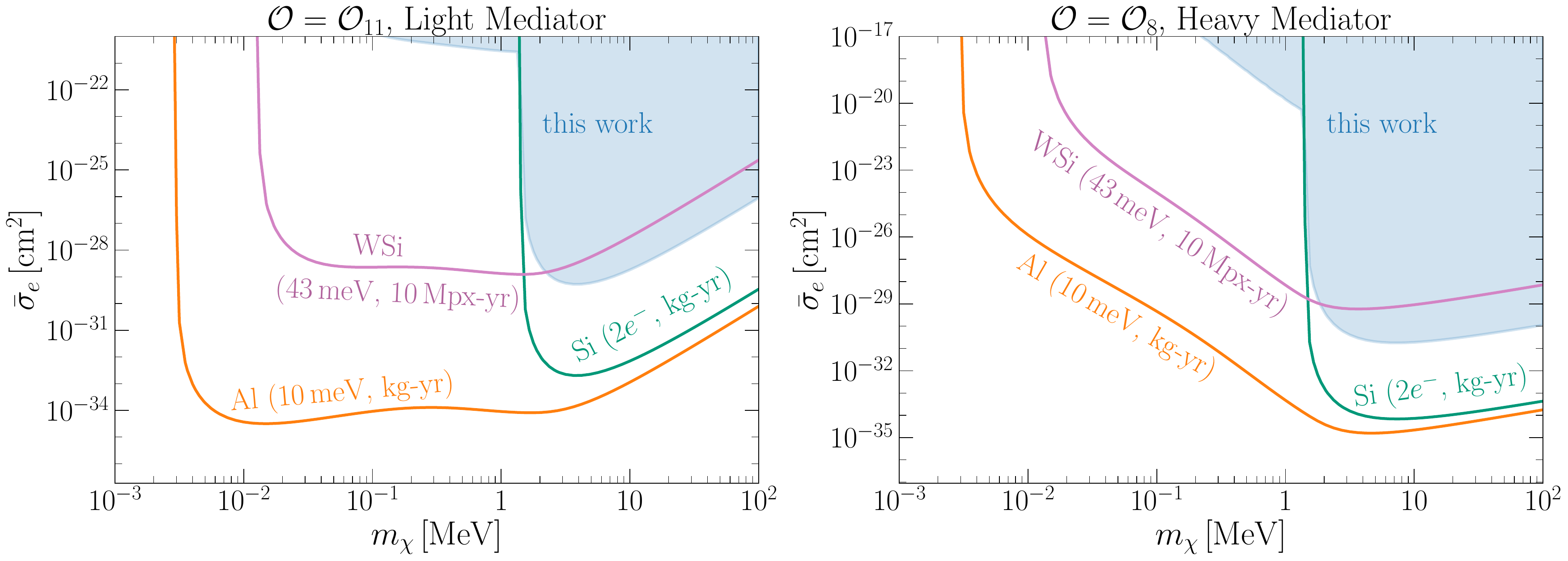}
    \caption{{\bf Spin-dependent results using the dielectric function.} Our 95\% C.L. new bounds and projections on  electric dipole DM ${\cal O} = {\cal O}_{11}$ with a light mediator~({\it left}) and anapole DM ${\cal O} = {\cal O}_8$ with a heavy mediator~({\it right}).
    The blue shaded regions~(labeled `this work') correspond to the new bounds we place using data from the QROCODILE~\cite{Baudis:2025zyn} and DAMIC-M~\cite{DAMIC-M:2025luv} collaborations.
    The solid curves correspond to future projections for several materials assuming three orders of magnitude sensitivity and no background events.
    The solid pink curve corresponds to $1$~yr exposure of $10^7$ WSi pixels, each of similar configuration to the current QROCODILE pixel~\cite{Baudis:2025zyn},  with a threshold of $42.8 \, {\rm meV}$.
    The green (orange) solid curves delineate projections for a kg-yr exposure of Si (superconducting Al) with a threshold of $1.12 \, {\rm eV}$ ($10 \, {\rm meV}$). 
    }
    \label{fig:dielectric spin-dependent bounds}
\end{figure*}
%%%%%%%%%%%%%%%%%%%%%%%%%%%%%%%%%%%%%%%%%%%%%%%%%%%%%%%

Eq.~\eqref{eq:master_eq} shows that the DM-electron interaction rate resulting from operators independent of the electron spin--- namely ${\cal O}_1, {\cal O}_{2}, {\cal O}_5, {\cal O}_8$ and ${\cal O}_{11}$---is completely determined by the dielectric response ${\rm Im}(-\chi_{00})$. 
Furthermore, for the operators ${\cal O}_5, {\cal O}_8$ and ${\cal O}_{11}$, the form-factor $\SIff$ in Eq.~\eqref{eq:form factors} depends on the \emph{DM spin}, thereby making it possible to determine the rate of such spin-dependent interactions (\textit{i.e.} any sum of ${\cal O}_{5},{\cal O}_{8}$ and ${\cal O}_{11}$) solely via the dielectric function. 
Here we demonstrate this property for the operators ${\cal O}_{11} = i \vb{S}_\chi \cdot \vb{q} / m_\chi$ of electric dipole moment DM and ${\cal O}_{8} = \vb{S}_\chi \cdot \vb{v}_\perp$ which arises {\it e.g.} in anapole DM \cite{Ho:2012bg}. (Note that the projections for anapole DM in Ref.~\cite{Trickle:2019ovy,Berlin:2025uka} focus on the {\it electron} spin-dependent operator ${\cal O}_9$, which is of similar order in velocity suppression to ${\cal O}_8$, but whose rate is determined via the spin response, not the dielectric tensor. 
Both ${\cal O}_8$ and ${\cal O}_9$ arise in the non-relativistic limit of the anapole DM model.)
For both operators ${\cal O}_8$ and ${\cal O}_{11}$, $\SIff = {\cal O}_{i}$ and $\SDff = 0$.
In particular, Eq.~\eqref{eq:master_eq} reads 
\begin{eqnarray}\label{eq:diff rate 1,8, 11}
    \frac{\dd \Gamma_{1}}{\dd^3 \vb{q}} & = & \frac{2 g_{e\chi}^2}{(q^2 + m_\phi^2)^2} \, {\rm Im}\Bqty{-\chi_{00}(\omega_{\vb{q}}, \vb{q})}  \, ,
    \newrow{}
    \frac{\dd \Gamma_{8}}{\dd^3 \vb{q}} & = & \frac{v_\perp^2}{4} \frac{\dd \Gamma_{1}}{\dd^3 \vb{q}} \, , 
    \qquad \frac{\dd \Gamma_{11}}{\dd^3 \vb{q}} = \frac{q^2}{4 m_\chi^2} \frac{\dd \Gamma_{1}}{\dd^3 \vb{q}} \, , 
\end{eqnarray}
where both differential rates are expressed in terms of the differential rate $\Gamma_1$ for the spin-independent benchmark operator ${\cal O}_1 = 1$. 
In this way, it becomes evident that one can recast existing spin-independent measurements into the complementary parameter space of these spin-dependent interactions.

We place the first constraints on the DM-electron interactions of an electric dipole ${\cal O}_{11}$ with a light mediator~({\it left}) and ${\cal O}_8$ with a heavy mediator~({\it right})  in Fig.~\ref{fig:dielectric spin-dependent bounds}, using data from the QROCODILE~\cite{Baudis:2025zyn} and DAMIC-M~\cite{DAMIC-M:2025luv} collaborations. (Note that in the conventions of this manuscript the choice of ${\cal O}_8$ with a heavy mediator mimics the ${\cal O}_8$ contribution of anapole DM mediated by a light vector.)
The QROCODILE experiment uses a WSi superconducting nanowire single photon detector~(SNSPD) pixel of mass $1.67 \, {\rm ng}$, and collected data over $415.15 \, {\rm hrs}$ with a detection threshold of $\sim 113 \, {\rm meV}$. 
We model the WSi dielectric response using a Lindhard function~\cite{Dressel_Gruner:2002} with plasmon frequency $\omega_p = 10.8 \, {\rm eV}$ and width $\Gamma_p = 0.7 \, {\rm eV}$.
For the DAMIC-M data, we perform a single bin analysis of the $2e^-$ excitations, corresponding to a detection threshold of $4.71\, {\rm eV}$ in their silicon skipper charge-coupled device (CCD), and model the Si dielectric response using a Lindhard function with plasmon frequency $\omega_p = 17.33\, {\rm eV}$ and width $\Gamma_p = 1.13 \, {\rm eV}$ found by fitting the {\tt Materials Project}~\cite{Jain:2013wst,Petousis:2017} absorption spectra.
This procedure for recasting the DAMIC-M data is accurate to the ${\cal O}(1)$ level and is expected to yield a conservative estimate. 
The combined new constraints we derive are shown in shaded blue. Solid colored curves indicate future projections for several interesting benchmarks. The reach of a next generation design of the QROCODILE experiment, with  an exposure of $10^7$ WSi SNSPD pixels for $1\, {\rm yr}$ with a detection threshold of $42.8 \, {\rm meV}$~\cite{2023Optic..10.1672T}, is indicated in solid purple. Future projections for Si and superconducting Al~\cite{Hochberg:2015pha,Hochberg:2015fth,Hochberg:2021pkt} with detection thresholds $1.12 \, {\rm eV}$ and $10\, {\rm meV}$, respectively, are shown in solid green and solid orange, respectively, assuming a background-free ${\rm kg}$-${\rm yr}$ exposure.
We model the Al response via a Lindhard function with a plasmon frequency $\omega_p = 15.8 \, {\rm eV}$ and plasmon width $\Gamma_p = 1.58 \, {\rm eV}$.
All bounds and projections are given at the 95\% C.L. and assume a dynamic range spanning three orders of magnitude in energy above threshold.
(We note that our projection for Si, that utilizes the charge density response (dielectric function) alone, yields similar reach within ${\cal O}(1)$ to the much more complicated treatment, involving multiple material responses, that appears in Ref.~\cite{Catena:2024rym}.)

As expected from Eq.~\eqref{eq:diff rate 1,8, 11}, our new constraints on the anapole and electric dipole DM interactions from existing SNSPD and skipper CCD data are roughly suppressed by the DM velocity $|\vb{v}|^2 \sim 10^{-6}$ compared to the constraints placed in the literature on the spin-independent ${\cal O}_1$ cross-section by the experimental collaborations.  
Since coherent spin-dependent effects are not predicted to be present in the atmosphere, we expect the upper bound from overburden to scale similarly.
Thereby, as long as the spin-independent bounds exceed overburden by several orders of magnitude, the velocity-suppressed spin-dependent bounds are expected to scale similarly. 
Recent works~\cite{DAMIC-M:2023hgj, Bertou:2025adb} demonstrate how overburden can contribute to the daily modulation of the signal.
A detailed investigation of the exact upper bound and the resulting directional signal from overburden for the different operators is left for future work.

%%%%%%%%%%%%%%%%%%%%%%%%%%%%%%%%%%%%%%%
\subsection{Electronic Spin Response}\label{ssec:spin}
%%%%%%%%%%%%%%%%%%%%%%%%%%%%%%%%%%%%%%%
%
%%%%%%%%%%%%%%%%%%%%%%%%%%%%%%%%%%%%%%%
\begin{figure*}
    \centering
\includegraphics[width=0.47\linewidth]{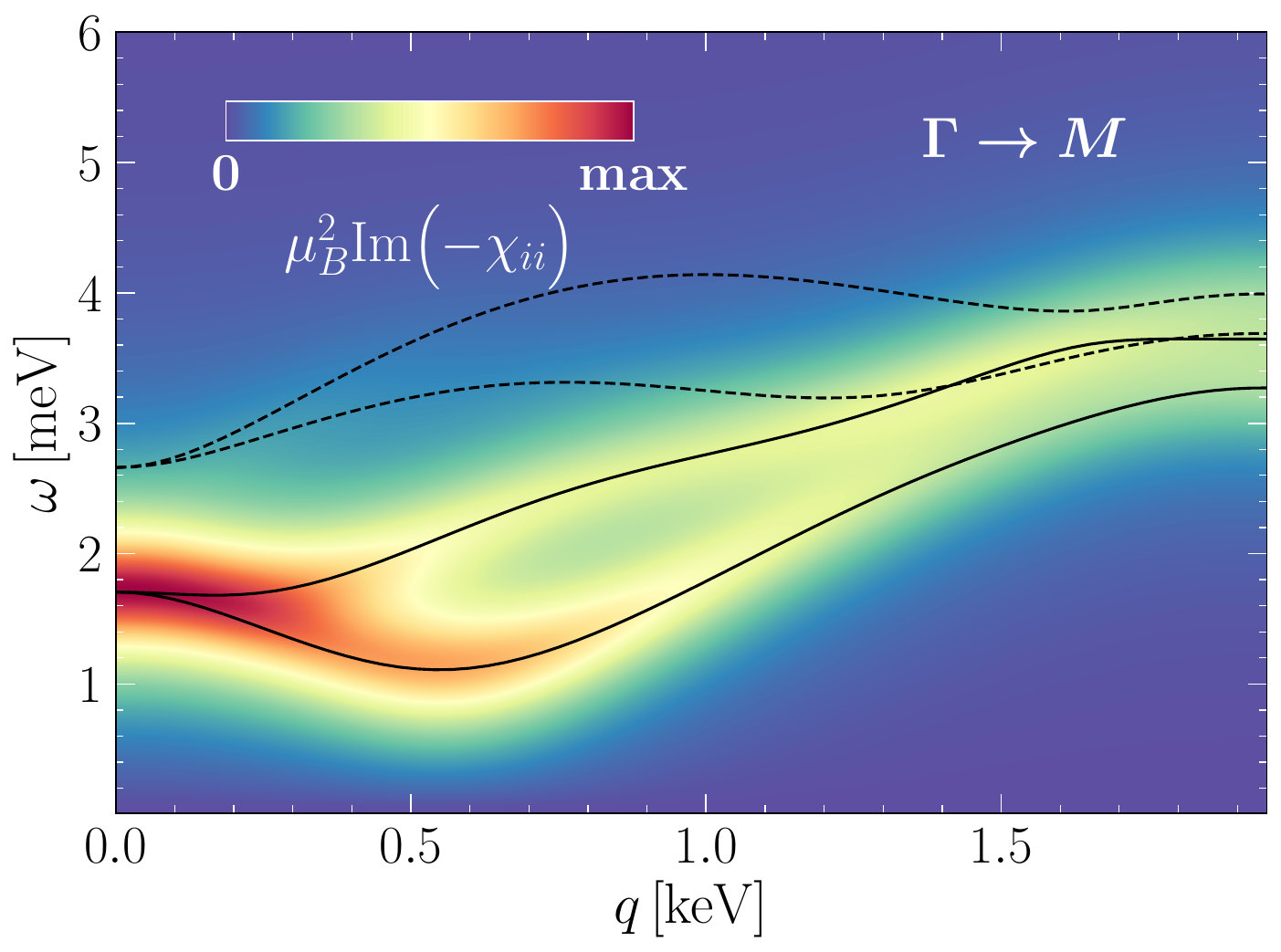}
\hspace{0.5cm}
\includegraphics[width=0.47\linewidth]{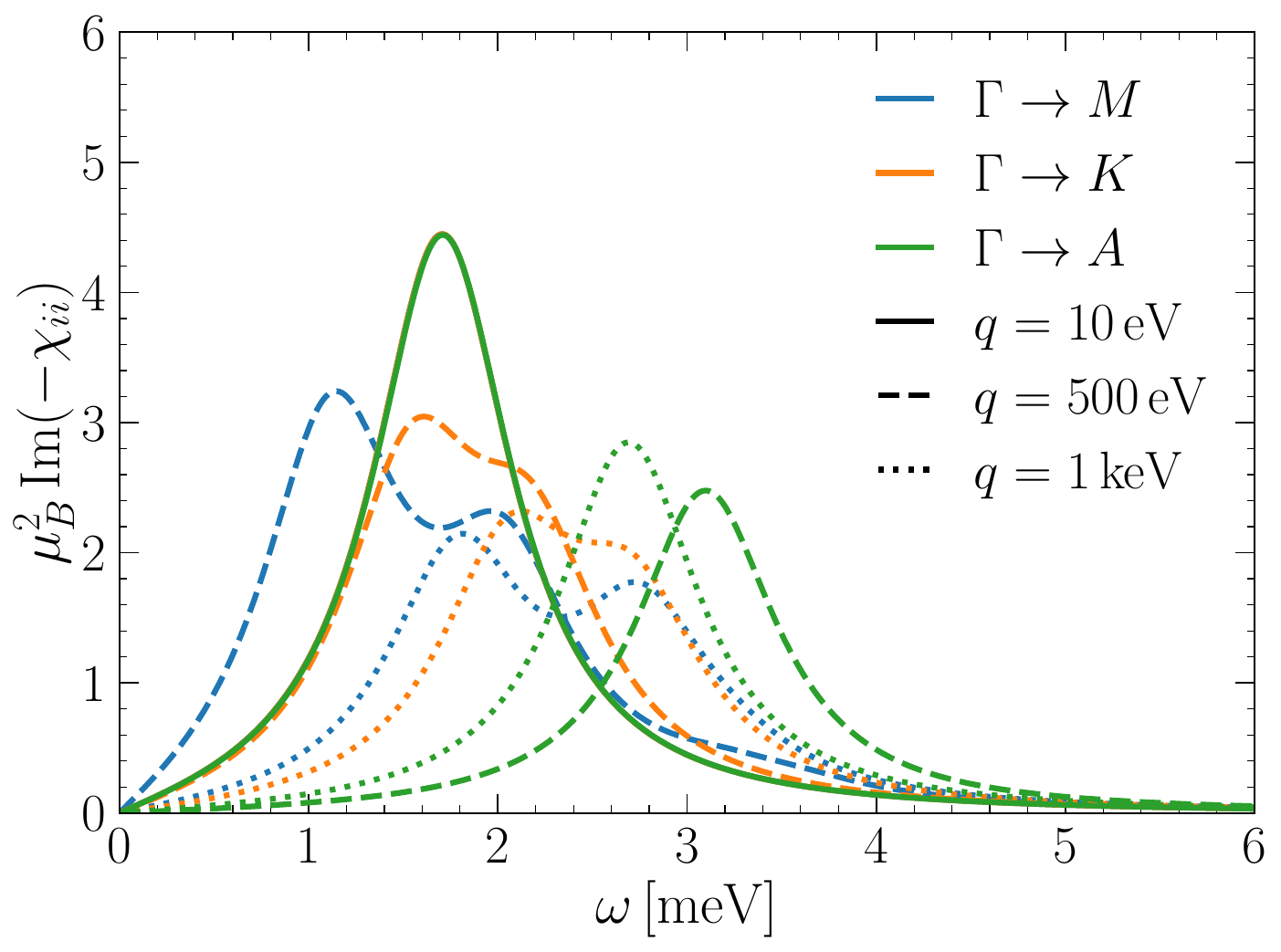}
    \caption{{\bf Pr spin response.} {\it Left.} The trace of the spin response of Pr for momenta in the $\Gamma \to M$ crystal direction. 
    The black lines indicate the dispersion relations of the different modes. 
    {\it Right.}~The trace of the spin response of Pr at momenta $q=10 \, {\rm eV},\, 500 \, {\rm eV},\, 1 \, {\rm keV}$ along the $\Gamma \to M, K, A$ crystal directions. A constant excitation width of $\Gamma=0.45$ meV has been used to calculate the response. We find the response depends on both the magnitude and direction of the momenta. 
    }
    \label{fig:pr_response}
\end{figure*}  
%%%%%%%%%%%%%%%%%%%%%%%%%%%%%%%%%%%%%%%

Eq.~\eqref{eq:master_eq} shows that the rate for interactions
between DM and the electron spins involves the electronic spin response $\chi_{ij}$.
Similarly to the dielectric function, it can be experimentally measured (most commonly through neutron scattering)~\cite{Boothroyd:2020, Berlin:2025uka}, or computed using analytical models and density functional theory (DFT)~\cite{Cao:2018dft,Buczek:2011dft,Rousseau:2012dft,Gorni:2018dft,Binci:2025dft}.
In this context, we consider the paramagnetic rare earth metal Pr which exhibits strong anisotropic responses at the meV scale due to the existence of a $\Delta \simeq 3.5 \, {\rm meV}$ energy gap between its ground state and the first two excited states.
In particular, we consider Pr atoms in a double hexagonal close-packed lattice (DHCP), where we use a simplified model for the multi-body response of the electrons in these three energy levels to evaluate the spin response~\cite{Jensen:1991, Boothroyd:2020}.
The model and simplifying assumptions are described in Appendix~\ref{app:Praseodymium}.
We note that this modeling tends to underestimate the crystal response since it neglects a variety of higher-energy excited states that are present in the crystal.
The model parameters are calibrated by matching to neutron scattering experiments~\cite{Houmann:1971, Bak:1975, Houmann:1979}, providing a good fit to the dispersion of the low-energy excitation spectrum~of~Pr.
The modeling further obeys the Kramers–Kronig relations, as required by causality.
The excitation width $\Gamma$ is  undetermined by the model, with measured values ranging from $0.45\, {\rm meV}$ to $\sim 2 \, {\rm meV}$~\cite{Houmann:1979}.
We further include the magnetic form factor of the Pr ions~\cite{Lebech1979}, which provides an isotropic momentum-dependent suppression, with the response vanishing at $q \gtrsim 2 \, {\rm keV}$.

We present the trace of the spin response  ${\rm Im} \pqty{-\chi_{ii}}$ for a Pr crystal in Fig.~\ref{fig:pr_response} as calculated from the analytical model with $\Gamma = 0.45\, {\rm meV}$. 
The left panel shows the spin response along a particular crystal direction as a function of the deposited energy and momentum. 
(The response along other crystal directions and additional information can be found in Fig.~\ref{fig:pr_response_appendix} of Appendix~\ref{app:Praseodymium}.) 
At low momenta, we identify a peaked response at energies of order a few meV, indicating sensitivity to keV-scale DM scattering.
The peaked response at keV-scale momenta extends the reach to MeV-scale DM masses.
The right panel of Fig.~\ref{fig:pr_response} presents the spin response along several different momenta directions at various fixed values of $q$ as a function of energy, demonstrating the anisotropy of the response. 
Additional anisotropy is manifest by the different responses along different spin directions $\chi_{11} \neq \chi_{22}$ (see  Appendix~\ref{app:Praseodymium}). 

%%%%%%%%%%%%%%%%%%%%%%%%%%%%%%%%%%%%%%%
\begin{figure*}
    \centering
    \includegraphics[width=0.98\linewidth]{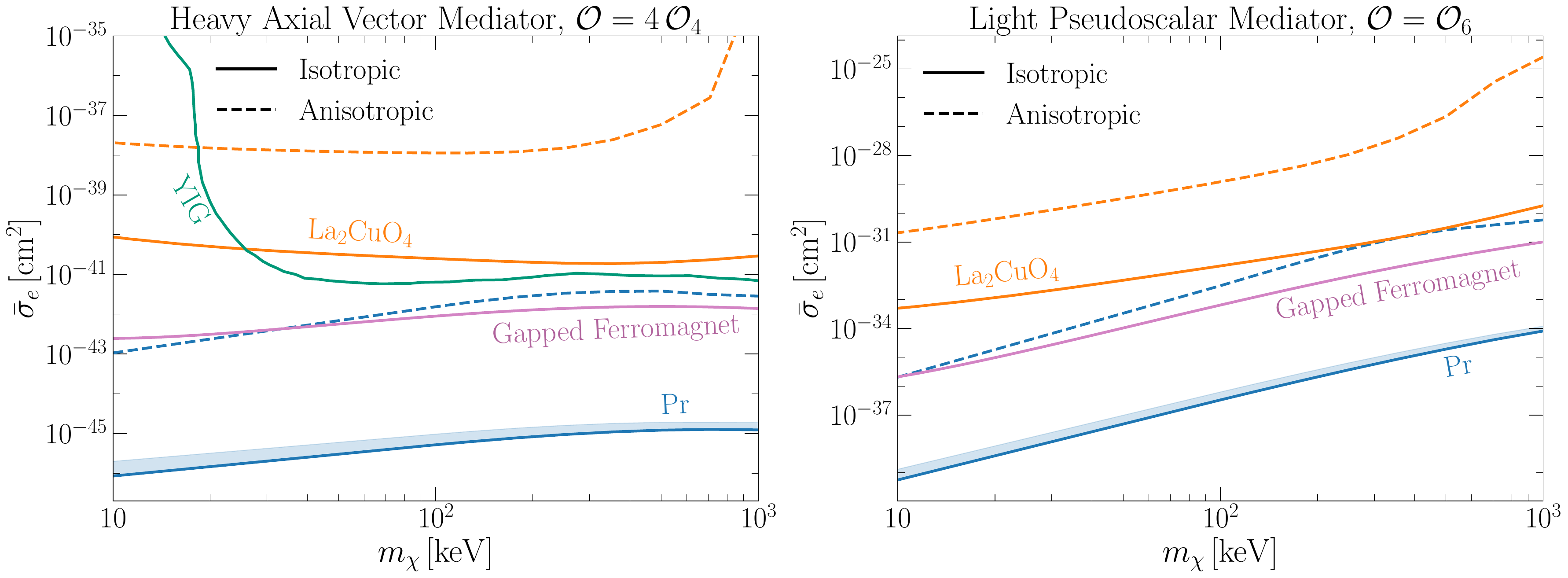}
    \caption{{\bf Spin-dependent results using spin response.} Projected reach for DM-electron interactions mediated by a heavy axial vector ${\cal O} = 4 {\cal O}_4$ ({\it left}) and light pseudoscalar ${\cal O} = {\cal O}_6$ ({\it right}) for Pr, where the shaded region indicates variation with the width $\Gamma$, the anti-ferromagnet La$_2$CuO$_4$ (orange, using the model of Ref.~\cite{Marocco:2025eqw}), and a gapped ferromagnet (pink, see text for  details). Curves are computed at 95\% C.L. for a kg-year exposure and energy acceptance $\omega \in [1\, {\rm meV}, 1 \, {\rm eV}]$. 
    {\it Solid} curves delineate the isotropic reach, whereas {\it dashed} curves delineate the  directional reach.
    In the left panel, we also show for comparison the projected reach of YIG (green) with a $25\,{\rm meV}$ threshold~\cite{Trickle:2020oki}.
    }
    \label{fig:spin response bounds}
\end{figure*}
%%%%%%%%%%%%%%%%%%%%%%%%%%%%%%%%%%%%%%%

As benchmark models we consider an axial vector and pseudoscalar (axion) mediator corresponding to the interactions ${\cal O} = 4 \, {\cal O}_4 = 4 \, \vb{S}_\chi \cdot \vb{S}_e$ and ${\cal O} = {\cal O}_6 = (\vb{q} \cdot \vb{S}_\chi)(\vb{q} \cdot \vb{S}_e) / (m_e m_\chi)$ respectively.
For both operators $F_0=0$; for the axial vector mediator $\vb{F} = 4 \, \vb{S}_\chi$, while for the pseudo-scalar mediator $\vb{F} = (\vb{q}\cdot \vb{S}_\chi) \vb{q} / (m_e m_\chi)$.
Using Eq.~\eqref{eq:master_eq}, the corresponding DM scattering rates are
\begin{eqnarray}
    \frac{\dd \Gamma_{4}}{\dd^3 \vb{q}} & = & \frac{8 g_{e\chi}^2}{(q^2 + m_\phi^2)^2} \, {\rm Im}\Bqty{-\chi_{ii}(\omega_{\vb{q}}, \vb{q})}
    \newrow{,}
    \frac{\dd \Gamma_{6}}{\dd^3 \vb{q}} & = & \frac{g_{e\chi}^2 \, q^2 q_i q_j}{2m_e^2 m_\chi^2 (q^2 + m_\phi^2)^2}\, {\rm Im}\Bqty{-\chi^+_{ij}(\omega_{\vb{q}}, \vb{q})} \, .
\end{eqnarray}
Similarly, the operators ${\cal O}_3, {\cal O}_7, {\cal O}_9, {\cal O}_{10}, {\cal O}_{12}, {\cal O}_{13}, {\cal O}_{14}$ and $ {\cal O}_{15}$ also exhibit $F_0=0$, meaning the DM interaction rate is entirely captured by the spin-response $\chi_{ij}$.

In Fig.~\ref{fig:spin response bounds} we present our projections for heavy axial vector mediated ({\it left}) and light pesudoscalar mediated ({\it right}) DM scattering off a Pr target with kg-yr exposure, assuming no backgrounds and an energy acceptance in the range $[1\,{\rm meV}, 1\,{\rm eV}]$. 
The blue shaded region indicates the variation of our projection for different widths $\Gamma \in [0.45 \, {\rm meV}, 2 \, {\rm meV}]$, with the solid curve corresponding to the smaller width.
For comparison, under the same assumptions, we show the reach of the anti-ferromagnet La$_2$CuO$_4$, with the response modeled according to the magnon model of Ref.~\cite{Marocco:2025eqw}, and an example of a ferromagnet with a gapped low-energy excitation, representing an optimistic benchmark for several materials studied in Refs.~\cite{Trickle:2019ovy,Trickle:2019nya,Marocco:2025eqw}. 
The gapped ferromagnet is assumed to have a spin response $\Im(-\chi^+_{ij}) = n_s \delta_{ij} \delta(\omega-\Delta)$ with a gap $\Delta = 5 \, {\rm meV}$, and spin density $n_s=5\times 10^{21}\, {\rm cm}^{-3}$; the material density is assumed to be $\rho_{\rm T} = 4\, {\rm g}\,{\rm cm}^{-3}$.
For the axial vector mediator model, we also show the projected reach of YIG~\cite{Trickle:2020oki} with a $25 \, {\rm meV}$ detection threshold for comparison, noting that it uses slightly different parameters for the DM velocity distribution.

The directional reach of Pr and La$_2$CuO$_4$ are delineated by the dashed curves in Fig.~\ref{fig:spin response bounds}.
We define the sensitivity to a directionally modulating signal by the minimal number of events required to distinguish a difference between the number of events observed in AM vs. PM at 95\% C.L. 
This two-bin probability is calculated using Monte Carlo, using the procedure described in Refs.~\cite{Griffin:2025wew, Hochberg:2025dom}.
Our analysis provides a conservative estimate; analyses utilizing more time bins are expected to improve the directional reach further.

Note that we use different crystal orientations to calculate the  isotropic and anisotropic (directional) reach.
Both Pr and La$_2$CuO$_4$ have much larger responses in the $xy$ plane than in the $\vu{z}$ direction, with the latter approximated to vanish in the analytical models. 
(This approximation slightly increases the directional signal, although the overall effect on the directional reach can only be determined by the use of a more precise material response incorporating all material excitations.)
As such, the largest modulation of a DM signal is achieved when the DM wind is perpendicular to the $xy$ plane, whereas the largest event count is achieved when the modulation is parallel to the plane.
To reduce computation time, here we have presented the best reach among the three cartesian directions: $\vu{z}$ for the isotropic reach and $\vu{x}$ for the directional reach; an optimal orientation can be found by sampling a dense grid of crystal orientations.

Our findings demonstrate that Pr provides a promising avenue for sensing spin-dependent DM-electron scattering down to the keV scale, for both isotropic and directional searches.
We also find that a ferromagnet with a typical spin density and a single gapped state at the meV scale outperforms existing proposals by an order of magnitude (see also Ref.~\cite{Trickle:2020oki} for the projected reach of the ferromagnet $\alpha$-RuO$_4$). 
Pr benefits from a relatively large average electron spin density and two highly degenerate gapped modes at the meV scale, boosting its sensitivity to DM by up to six orders of magnitude compared to current proposals.
(Note that the bound presented in Ref.~\cite{Catena:2025sxz} assumes $\bqty{-\chi_{ij}(\omega=0, \vb{q})} < q^2 / (2\pi \alpha)$, which does not apply to our case.  
A more general upper bound which applies to all material responses is the subject of upcoming work~\cite{bound-to-appear}.)

%%%%%%%%%%%%%%%%%%%%%%%%%%%%%%%%%%%%%%
\section{Outlook}\label{sec:sum}
%%%%%%%%%%%%%%%%%%%%%%%%%%%%%%%%%%%%%%
%
Laboratory searches for sub-MeV DM are increasingly focused on DM-electron couplings thanks to their favorable kinematics and their potential sensitivity to sub-eV energy deposits realized through collective excitations in various target materials. 
Indeed, such targets can also probe DM-nuclear couplings at sub-MeV scales~\cite{Griffin:2024cew}. 
The choice of materials and their specific properties play a crucial role in determining detection capabilities. 
Our master equation Eq.~\eqref{eq:master_eq} serves as a bridge between physical material properties---three types of measurable material responses---and the prospects for light DM detection, extending the seminal works of Refs.~\cite{Hochberg:2021pkt, Knapen:2021bwg, Boyd:2022tcn} to all DM-electron interactions.

Our presentation makes clear that, in contrast to common lore, even for \textit{spin-dependent} interactions---as long as these depend only on the spin of the DM and not on the electron spin---the relevant material response that governs the DM interaction rate is fully described by the dielectric tensor.
Leveraging this result, we derive the first direct detection constraints on anapole and electric dipole DM using data from the QROCODILE~\cite{Baudis:2025zyn} and DAMIC-M~\cite{DAMIC-M:2025luv} experiments, and provide projections for future experimental searches.

Additionally, we have demonstrated how to use the material spin-response $\chi_{ij}$ to find the projected reach for spin-dependent DM-electron interactions.
We have shown that commercially available Pr crystals offer prospects---both for isotropic scattering and directional detection---that surpass existing proposals~\cite{Trickle:2019ovy, Trickle:2020oki, Marocco:2025eqw, Berlin:2025uka} by several orders of magnitude.
Our findings suggest that Pr, and possibly other rare earth metals, offer a promising avenue for probing spin-dependent interactions.

Although the charge-spin response $\chi_{0i}$ was not explicitly analyzed in this work, it points to an exciting opportunity: leveraging the mature and rapidly advancing technology of spintronics for DM detection. 
Spintronics exploit the charge-spin response, often generated via spin-orbit coupling, to manipulate and detect spin using charge-based measurements.  
In addition to expanding the breadth of materials one may consider for DM detection, it also introduces a novel detection philosophy, where DM-induced spin polarization can be read out through electrical resistance. 
Several high-throughput material searches are currently targeting large spin Hall conductivities~\cite{Meinert:2020,Zhang:2021}, which are directly related to $\chi_{0i}$. 
These efforts could be capitalized on by using an approach similar to that of Ref.~\cite{Griffin:2025wew}. 
We leave a detailed investigation of these directions to future work.

Our formalism bridges the gap between material science,  experimental spectroscopy and DM direct detection, allowing the DM detection community to harness existing experimental efforts and to forge new pathways for DM detection.

\bigskip
%%%%%%%%%%%%%%%%%%%%%%%%%%%%%%%%%%%%%%%%%%%%%%%%%%%%%%%%%%%%%%%%%%
\textbf{Acknowledgments.} 
We thank Dror Orgad, Zohar Komargodski and Zohar Ringel for useful discussions, and Ben Lehmann and Dino Novko for valuable comments on the manuscript. 
The work of Y.H. is supported by the Israel Science Foundation (grant No. 1818/22) and by the Binational Science Foundation (grants No. 2018140 and No. 2022287). Y.H., M.K., A.L. and R.O. are supported by an ERC STG grant (``Light-Dark,'' grant No. 101040019). 
M.K., A.L. and R.O. are grateful to the Azrieli Foundation for the award of an Azrieli Fellowship. 
The work of M.K. is also supported by the BSF grant No.
2020220. R.O. is also supported by BSF Travel Grant No. 3083000028 and the Milner Fellowship. 
Y.H, M.K., A.L and R.O. thank Cornell University for their gracious hospitality.  
This project has received funding from the European Research Council (ERC) under the European Union’s Horizon Europe research and innovation programme (grant agreement No. 101040019).  Views and opinions expressed are however those of the author(s) only and do not necessarily reflect those of the European Union. The European Union cannot be held responsible for them. 
%%%%%%%%%%%%%%%%%%%%%%%%%%%%%%%%%%%%%%%%%%%%%%%%%%%%%%%%%%%%%%%%%%

\appendix
\onecolumngrid

%%%%%%%%%%%%%%%%%%%%%%%%%%%%%%%%%%%%%%%%%%%%%%%%%%%%%%%%%%%%%%%%%%%%%%%%%%%%%
\section{Relation to Relativistic four-Fermi Interactions}\label{app:operator NR limit}
%%%%%%%%%%%%%%%%%%%%%%%%%%%%%%%%%%%%%%%%%%%%%%%%%%%%%%%%%%%%%%%%%%%%%%%%%%%%%
%
\begin{table*}\label{tab:interactions}
\centering
\scriptsize
\begin{tabular}{|cc|c|c|c|}
    \hline
    {\bf Interaction}  & & {\bf Model} & {\bf LO Non-Relativistic~Limit} & $\sum_i c_i {\cal O}_i$ \\
    \hline \hline
    $\bar{\chi} \chi \bar{e} e$ &   & Scalar Mediator & $1$ & $ \mathcal{O}_1$ \\[5pt]
    $i \bar{\chi} \chi \bar{e} \gamma^5 e$ &  &  & $ i\dfrac{\vb{q}}{m_e} \cdot \vb{S}_e$ & $ \mathcal{O}_{10}$ \\
    $i \bar{\chi} \gamma^5 \chi \bar{e} e$ & & & $ - i\dfrac{\vb{q}}{m_\chi} \cdot \vb{S}_\chi$ & $ -\mathcal{O}_{11}$ \\
    $\bar{\chi} \gamma^5 \chi \bar{e} \gamma^5 e$ &   & Pseudoscalar Mediator & $- \dfrac{  \vb{q}}{m_\chi} \cdot \vb{S}_\chi \dfrac{\vb{q}}{m_e} \cdot \vb{S}_e$ & $-  \mathcal{O}_6$ \\
    $\dfrac{P^\mu}{2m_\chi} \bar{\chi} \chi \dfrac{K_\mu}{2m_e} \bar{e} e$ &  & Vector Mediator & $1$ & $  \mathcal{O}_1$ \\
    $\dfrac{P^\mu}{2m_\chi} \bar{\chi} \chi \bar{e} i \sigma_{\mu \alpha} \dfrac{q^\alpha}{2m_e} e$ &  &  & $- \dfrac{q^2}{4m_e^2} -i   \vb{v}_\perp \cdot\left(\dfrac{\vb{q}}{m_e} \times \vb{S}_e\right)$ & $- \dfrac{q^2}{4m_e^2}\mathcal{O}_1+ \mathcal{O}_3$ \\
    $\dfrac{P^\mu}{2m_\chi} \bar{\chi} \chi \bar{e} \gamma_\mu \gamma^5 e$ & & & $-2 \,  \vb{v}_\perp \cdot \vb{S}_e$ & $-2 \, \mathcal{O}_7$ \\
    $i \dfrac{P^\mu}{2m_\chi} \bar{\chi} \chi \dfrac{K_\mu}{2m_e} \bar{e} \gamma^5 e$ & & & $  i \dfrac{\vb{q}}{m_e} \cdot \vb{S}_e$ & $ \mathcal{O}_{10}$ \\
    $\bar{\chi} i \sigma^{\mu \nu} \dfrac{q_\nu}{2m_\chi} \chi \dfrac{K_\mu}{2m_e} \bar{e} e$ &   & Magnetic Dipole DM & $ \dfrac{q^2}{4m_\chi^2} + i  \vb{v}_\perp \cdot\left(\dfrac{\vb{q}}{m_\chi} \times \vb{S}_\chi\right)$ & 
    $\dfrac{{q^2}}{4m_\chi^2}\mathcal{O}_1-  \mathcal{O}_5$ \\
    $\bar{\chi} i \sigma^{\mu \nu} \dfrac{q_\nu}{2m_\chi} \chi \bar{e} i \sigma_{\mu \alpha} \dfrac{q^\alpha}{2m_e} e$ &   & Magnetic Dipole DM & $ \left(\dfrac{\vb{q}}{m_\chi} \times \vb{S}_\chi\right) \cdot\left(\dfrac{\vb{q}}{m_e} \times \vb{S}_e\right)$ & $ \dfrac{q^2}{m_\chi m_e} \mathcal{O}_4-\mathcal{O}_6 $ \\
    $\bar{\chi} i \sigma^{\mu \nu} \dfrac{q_\nu}{2m_\chi} \chi \bar{e} \gamma^\mu \gamma^5 e$ 
    &  &  & $-2i \, \vb{S}_e \cdot\left(\dfrac{\vb{q}}{m_\chi} \times \vb{S}_\chi\right)$ & $-2\dfrac{m_e}{m_\chi} \mathcal{O}_9$\\
    $i\bar{\chi} i \sigma^{\mu \nu} \dfrac{q_\nu}{2m_\chi} \chi \dfrac{K_\mu}{2m_e} \bar{e} \gamma^5 e$ &   &  & $ \left(i\dfrac{q^2}{4m_\chi^2} - \vb{v}_\perp \cdot\left(\dfrac{\vb{q}}{m_\chi} \times \vb{S}_\chi\right)\right) \dfrac{\vb{q}}{m_e} \cdot \vb{S}_e$ & 
    $\dfrac{q^2}{4 m_\chi^2} \mathcal{O}_{10} + \dfrac{q^2}{m_e m_\chi}{\cal O}_{12} + {\cal O}_{15}$ \\
    $\bar{\chi} \gamma^\mu \gamma^5 \chi \dfrac{K_\mu}{2m_e} \bar{e} e$ &   & Anapole DM & $2 \vb{v}_\perp \cdot \vb{S}_\chi$ & $ 2  \, \mathcal{O}_8$ \\
    $\bar{\chi} \gamma^\mu \gamma^5 \chi \bar{e} i \sigma_{\mu \alpha} \dfrac{q^\alpha}{2m_e} e$  &    & Anapole DM & 
    $-2i \,\vb{S}_\chi \cdot\left(\dfrac{\vb{q}}{m_e} \times \vb{S}_e\right)$ & $2 \, \mathcal{O}_9$ \\
    $\bar{\chi} \gamma^\mu \gamma^5 \chi \bar{e} \gamma^\mu \gamma^5 e$ &   & Axial Vector Mediator & $- 4\vb{S}_\chi \cdot \vb{S}_e$ & $- 4 \, \mathcal{O}_4$ \\
    $i \bar{\chi} \gamma^\mu \gamma^5 \chi K^\mu \bar{e} \gamma^5 e$ &   &  & $2i \, \pqty{\vb{v}_\perp \cdot \vb{S}_\chi}  \pqty{\dfrac{\vb{q}}{m_e} \cdot \vb{S}_e}$ & $2\, \mathcal{O}_{13} $ \\
    $i \dfrac{P^\mu}{2m_\chi} \bar{\chi} \gamma^5 \chi \dfrac{K_\mu}{2m_e} \bar{e} e$ &   & Electric Dipole DM & $-i \dfrac{\vb{q}}{m_\chi} \cdot \vb{S}_\chi$ & $-   \mathcal{O}_{11}$\\
    $i \dfrac{P^\mu}{2m_\chi} \bar{\chi} \gamma^5 \chi \bar{e} i \sigma_{\mu \alpha} \dfrac{q^\alpha}{2m_e} e$ &   & Electric Dipole DM  & 
    $ \dfrac{\vb{q}}{m_\chi} \cdot \vb{S}_\chi \left(i\dfrac{q^2}{4m_e^2}- i \vb{v}_\perp \cdot\left(\dfrac{\vb{q}}{m_e} \times \vb{S}_e\right)\right)$ 
    & $ \dfrac{q^2}{4m_e^2} \mathcal{O}_{11}  +  \mathcal{O}_{15}$ \\
    $i \dfrac{P^\mu}{2m_\chi} \bar{\chi} \gamma^5 \chi \bar{e} \gamma_\mu \gamma^5 e$ &   & 
    & $2i \pqty{\frac{\vb{q}}{m_\chi} \cdot \vb{S}_\chi} \pqty{\vb{v}_\perp \cdot \vb{S}_e}$ & $2 \,  \mathcal{O}_{14} $ \\
    $\dfrac{P^\mu}{2m_\chi} \bar{\chi} \gamma^5 \chi \dfrac{K_\mu}{2m_e} \bar{e} \gamma^5 e$ &   & Pseudoscalar Mediator & $-  \dfrac{\vb{q}}{m_\chi} \cdot \vb{S}_\chi \dfrac{\vb{q}}{m_e} \cdot \vb{S}_e$ & $ -\mathcal{O}_6$ \\
    \hline
\end{tabular}
\caption{{\bf Interaction mapping.}
Reduction of relativistic operators to non-relativistic effective operators which can be used with our master equation~\eqref{eq:master_eq}, similar to the familiar reductions done for DM-nucleon scattering in Refs.~\cite{Fitzpatrick:2012ix, Anand:2013yka, Gresham:2014vja}.
We denote $P^\mu \equiv p^\mu + p^{\prime \, \mu}$ where $p$ ($p'$) is the 4-momentum of the incoming (outgoing) DM and $K^\mu  \equiv k^\mu + k^{\prime \, \mu}$ where $k$~($k^\prime$) is the 4-momentum of the incoming (outgoing) electron. 
The second column indicates if the interaction corresponds to a particular model of interest following the naming conventions of Ref.~\cite{Gresham:2014vja}. 
The third column indicates the leading order~(LO) non-relativistic limit and the fourth column expresses the interaction in terms of the effective non-relativistic operators of Table~\ref{tab:NR_operators}.
Note that in the third and fourth columns, we have factored out $(4 m_e m_\chi)$ to match our choice of normalization for the non-relativistic wave function.
}
\label{table:DM-electron relativistic operators}
\end{table*}
Consider the interaction between fermionic DM and electrons. 
If this interaction is mediated by a heavy boson, we can integrate it out to obtain an effective description of the DM-electron interaction, resulting in four-Fermi operators.
Such dimension six operators in their non-relativistic form have become a common benchmark for DM direct detection~\cite{Fitzpatrick:2010br,Fitzpatrick:2012ix,Anand:2013yka,Gresham:2014vja}. 
Below, we provide an explicit example of how to derive the non-relativistic Lagrangian for a vector-mediated interaction between DM and electrons. 
This example serves two purposes:
it demonstrates how any interaction in the non-relativistic limit can be expressed as a combination of the operators in Table~\ref{tab:NR_operators}; and it motivates the definition of the function $V(q)$ in the examples shown in the main text. In particular, it justifies the use of $V(q)\propto q^{-2}$ for light mediators that cannot be simply integrated out to yield a local four-Fermi interaction.

We start from the interaction Lagrangian
\begin{align}
    {\cal L}  \supset g_{eA} \overline e \gamma^\mu A_\mu e  + g_{e\chi} \overline \chi \gamma^\mu A_\mu \chi\,,
\end{align}
where $A_\mu$ is a vector mediator of mass $m_A$ and $\chi$ is the fermionic DM. The matrix element for the elastic scattering process $\chi(p) e(k) \to \chi(p') e(k')$ is given by 
\begin{align}
    i{\cal M}  = \frac{-i\pqty{g_{\mu\nu} - \frac{q_\mu q_\nu}{q^2}}}{q^2 +m_A^2}(ig_{eA})(ig_{e\chi})[\overline u(k') \gamma^\mu u(k)][\overline u(p') \gamma^\nu u(p)]\ .
\end{align}
Since all the fermion legs are on-shell, the $q_\mu q_\nu$ term in the propagator vanishes, and one can use the Gordon identities to decompose the spinor bilinears into a vector monopole component and a magnetic dipole moment contribution,
\begin{align}\label{eq:gordon id}
    \overline u (p')\gamma^\mu u(p) = \overline u (p') \left( \frac{(p+p')^\mu}{2m} + \frac{i \sigma^{\mu\nu}(p'-p)_\nu}{2m}\right)u(p)\,,
\end{align}
with $\sigma^{\mu\nu} = i[\gamma^\mu, \gamma^\nu]/2$ and $m$ the mass of the fermion.
A complementary identity exists for an axial fermionic bilinear. 
Using the Gordon identity Eq.~\eqref{eq:gordon id}, the matrix element  decomposes into four terms:
\begin{align}
    i{\cal M}  = \frac{-ig_{\mu\nu}}{q^2 +m_A^2}\frac{(ig_{eA})}{2m_e}\frac{(ig_{e\chi})}{2m_\chi}\bigg\{ &[\overline u(k') K^\mu u(k)][\overline u(p') P^\nu u(p)]
    - [\overline u(k') K^\mu u(k)][\overline u(p') (i\sigma^{\nu\rho}q_\rho) u(p) ]\\\nonumber
    &+[\overline u(k') (i\sigma^{\mu\rho}q_\rho) u(k)][\overline u(p') P^\nu u(p)]
    - [\overline u(k') (i\sigma^{\mu\rho}q_\rho) u(k)][\overline u(p') (i\sigma^{\nu\rho}q_\rho) u(p) ]
    \bigg\}\ ,
\end{align}
where $K^\mu \equiv (k + k')^\mu$, $P^\mu \equiv (p + p')^\mu$ and $q^\mu = (k' - k)^\mu = (p - p')^\mu$. 
To obtain the non-relativistic limit, we can take the kinematics of DM scattering with a static electron in the lab frame: $p = (m_\chi + m_\chi v^2/2, m_\chi \vb{v})$, $p' = (m_\chi + m_\chi v^2/2 - \omega_{\vb{q}}, m_\chi \vb{v} - \vb{q})$, $k=(m_e, {\bf 0})$, $k'=(m_e+\omega_{\vb{q}}, {\bf q})$ and $\omega_{\vb{q}} = {\bf q}\cdot {\bf v} -q^2/(2m_\chi)$ and use the 4-spinors
\begin{eqnarray}
    u_s(p) = \sqrt{ p_0 + m} \pmqty{\xi_s \\ \frac{1}{p_0 + m } \vb*{\sigma} \cdot \vb{p} \, \xi_s} \, , \qquad \xi_s \in \Bqty{ \pmqty{1 \\ 0}, \pmqty{0 \\ 1}} \, , \qquad \bar{u}_s(p)u_{s'}(p) = 2m \delta_{ss'}\, ,
\end{eqnarray}
with the $\gamma^\mu$ matrices in the Dirac basis. The scattering amplitude up to next to leading order in the non-relativistic expansion is given by
\begin{eqnarray}\label{eq:scattering amplitide example}
    \frac{{\cal M}}{4m_e m_\chi} &=& \,\frac{g_{eA}\,g_{\chi A}}{q^2 +m_A^2} \Bigg\{ 
    \pqty{1 + \frac{q^2}{4m_\chi^2} + \frac{q^2}{4 m_e^2} + \frac{v_\perp^2}{2}} 
    - \pqty{ \frac{q^2}{4m_\chi^2} + i{\bf v}_\perp \cdot \left( \frac{\bf q}{m_\chi}\times {\bf S} _\chi\right)} \nn \\[5pt]
    &&\qquad\qquad\qquad
    - \pqty{ \frac{q^2}{4m_e^2} + i{\bf v}_\perp \cdot \pqty{ \frac{\bf q}{m_e}\times {\bf S} _e}}
    - \pqty{ \frac{\bf q}{m_e}\times {\bf S}_e } \cdot\pqty{ \frac{\bf q}{m_\chi}\times {\bf S}_\chi} 
    + \ldots
    \Bigg\}\,,
\end{eqnarray}
where the factors of $m_e$ and $m_\chi$ on the LHS indicate the normalization for the non-relativistic wave functions.
From Eq.~\eqref{eq:scattering amplitide example} we can read off the interaction Hamiltonian density in momentum space,
\begin{eqnarray}\label{eq:vector mediator hamiltonian}
    {\cal H}_{\rm int}(\vb{q}) = \frac{g_{eA} g_{\chi A}}{q^2 + m_A^2} 
    \Bqty{
    {\cal O}_1 + \frac{1}{2} {\cal O}_2 
    + {\cal O}_3 + {\cal O}_5 - \frac{q^2}{m_\chi m_e} {\cal O}_4 + {\cal O}_6
    } \, ,
\end{eqnarray}
from which we can identify 
\begin{eqnarray}
    \generalV = \frac{g_{eA} g_{\chi A}}{q^2 + m_A^2} \, , \qquad \SIff = {\cal O}_1 + \frac{1}{2} {\cal O}_2 - {\cal O}_5 \, , \qquad \SDff = i \frac{\vb{q}}{m_e} \times \vb{v}_\perp - \frac{q^2}{m_\chi m_e} \vb{S}_\chi + \pqty{\vb{S}_\chi \cdot \frac{\vb{q}}{m_\chi}} \, \frac{\vb{q}}{m_e} \, ,
\end{eqnarray}
in the notation of Eq.~\eqref{eq:Hint}. 
We emphasize that here we have kept sub-leading terms of order the velocity squared in the non-relativistic expansion for clarity. 
The leading order contribution is encapsulated entirely by the ${\cal O}_1$ term, {\it i.e.} $\SIff = 1$ and $\SDff = 0$.

The procedure above can be generalized, for {\it e.g.} scalar and pseudo-scalar mediators. 
For completeness in Table~\ref{table:DM-electron relativistic operators} we list the leading order non-relativistic limit of a variety of four-Fermi interactions, expressed in terms of both Galilean invariants and the effective operator basis listed in Table~\ref{tab:NR_operators}. 
Table~\ref{table:DM-electron relativistic operators} closely follows the structure of familiar tables for DM-nucleon interactions, {\it e.g.} the ones presented in Refs.~\cite{Fitzpatrick:2012ix, Anand:2013yka, Gresham:2014vja}, albeit with appropriate normalization for use with our master equation~\eqref{eq:master_eq}.
Additional relativistic operators can be composed as a sum of several relativistic operators in Table~\ref{table:DM-electron relativistic operators} via the Gordon identities.

%%%%%%%%%%%%%%%%%%%%%%%%%%%%%%%%%%%%%%%%%%%%%%
\section{Linear Response Theory}\label{app:linear response theory}
%%%%%%%%%%%%%%%%%%%%%%%%%%%%%%%%%%%%%%%%%%%%%%
%

Throughout this Appendix and Appendix~\ref{app:master eq derivation}, we carefully distinguish quantum mechanical operators from $c$-numbers using a hat, for clarity.

Consider a system consisting of a target ({\it e.g.} electrons in the detector) interacting with a probe ({\it e.g.} a DM particle) described by the Hamiltonian
\begin{align}
     \hat H = \hat H_0 + \hat{H}_{\rm int}\,.
\end{align}
We write the interaction Hamiltonian as 
\begin{align} \label{eqApp:linresp,Hint}
    \hat H_{\rm int} = 
    \int \dd^3 {\bf x}' \hat B_i ({\bf x}') \CF_i (\hat {\bf R}_\chi -{\bf x}')\,,
\end{align}
where $\hat B_i$ are operators representing the internal degrees of freedom of the target, $\CF_i$ are external forces acting on them, and $\hat {\bf R}_\chi$ corresponds to the probe's position. We assume the forces are turned on adiabatically and that $\hat H_{\rm int}$ is time-independent in the Schr\"odinger picture.
We denote asymptotic states for the system and the probe as $|s\rangle = |\Psi\rangle \otimes |p_s,\alpha_s\rangle$, assuming them to be factorized as is appropriate in the scattering limit.
Here $s = \{{\rm in},\  {\rm out}\}$, $\Psi$ denotes the state of the system, $p$ the four-momenta of the probe and $\alpha$ its internal degrees of freedom. 
We work in the interaction picture, where all the operators evolve with the free Hamiltonian $\hat H_0$, and states evolve with $\hat H_{\rm int}$.

In the weak probe limit $\langle \hat H_{\rm int} \rangle \ll \langle \hat H_0 \rangle$, the physical quantity describing the target's response is the susceptibility.
Given two target operators $\hat{A}(\vb{x})$ and $\hat{B}(\beta)$, where $\beta$ are internal indices and parameters on which $B$ depends ({\it{e.g.}} the spatial Fourier parameter $\vb{q}$), we define their correlation function~\cite{Boothroyd:2020} as
\begin{align} \label{eqApp:linresp, Dcorr}
    {\cal D}_{A B}(\tau,\vb{x}; \beta) \equiv -i\theta(\tau) \langle \Psi | \Big[\hat A(\tau,\vb{x}),\hat   B(\beta)\Big]| \Psi\rangle \,,
\end{align}
along with its Fourier transform, the \textit{generalized susceptibility},
\begin{align}
    \chi_{A B}(\omega,\vb{x};\beta) \equiv \int \dd \tau\, {\cal D}_{A B}(\tau,\vb{x};\beta)\,e^{i\omega \tau}\,.
\end{align}

Consider the fluctuations in some target observable $\hat A (t,\vb{x})$, 
\begin{align} \label{eqApp:linresp,DA}
    \langle \Delta \hat A \rangle (t,\vb{x}) \equiv \langle {\rm out} | \hat U^\dagger \hat A (t,\vb{x})  \hat U |{\rm in}\rangle - \langle {\rm out} | \hat A (t,\vb{x}) |{\rm in}\rangle\,, 
\end{align}
where $\hat U(t) = {\cal T} \exp\pqty{-i \int^t_{-\infty} \dd \tau\, \hat H_{\rm int}(\tau)}$ is the unitary time evolution operator.
Assuming a weak interaction $\langle\hat H_{\rm int}\rangle \ll \langle \hat H_0 \rangle$, as is appropriate in the DM-electron scattering scenario, we calculate the fluctuation Eq.~\eqref{eqApp:linresp,DA} to leading order in perturbation theory.
On the one hand, we can write 
\begin{align} \label{eqApp:linresp, DeltaAinChi}
    \langle \Delta \hat{A}\rangle (t,\vb{x}; -\vb{q}) = \int \dd t'\,\, {\cal D}_{A \langle H_{\rm int} \rangle}\left(t-t',\vb{x};-{\bf q}\right)\,e^{-i \Delta E t'} 
    = e^{-i \Delta E t}\, \chi_{A \langle H_{\rm int}\rangle}(\Delta E,\vb{x};-{\bf q})\,,
\end{align}
where $(\Delta E,{\bf q}) \equiv (E_{\rm in},{\bf p}_{\rm in})- (E_{\rm out},{\bf p}_{\rm out})$ are the energy and momentum transfer of the probe, and we defined
\begin{align}
    \langle \hat{H}_{\rm int}  \rangle (-\vb{q})\equiv \langle p_{\rm{out}},\alpha_{\rm{out}} |\hat{H}_{\rm int}| p_{\rm in},\alpha_{\rm in}\rangle=\langle \alpha_{\rm{out}} |\hat{\cal H}_{\rm int}(-\vb{q})|\alpha_{\rm in}\rangle\,.
\end{align}
Here, $\hat {\cal H}_{\rm int}(\vb{q})$ is the spatial Fourier transform of $\hat{\cal H}_{\rm int}$, in accordance with the convention that the spatial Fourier transform of a function $g(\vb{x})$ is given by $g(\vb{q})\equiv \int \dd^3\vb{x}\,\,g(\vb{x}) e^{-i \vb{q}\cdot\vb{x}}$.
Fourier-transforming Eq.~\eqref{eqApp:linresp, DeltaAinChi} with respect to $\vb{x},\,t$ yields 
\begin{align} \label{eqApp:linsresp,DAinChiAHint}
    \langle \Delta \hat A \rangle(\omega, {\bf q};-\vb{q}) = (2\pi) \delta (\omega - \Delta E)\,\chi_{A \langle H_{\rm int}\rangle}(\omega, {\bf q};-{\bf q})\,.
\end{align}
On the other hand, by using Eq.~\eqref{eqApp:linresp,Hint}, we obtain
\begin{align}\nonumber
    \langle \Delta \hat{A} \rangle  (t,\vb{x};-\vb{q}) &=  -i  \int_{-\infty}^{t} \dd t'\int \dd ^3 x'\bigg\{ 
     \langle \Psi | \Big[\hat A(t,\vb{x}), \hat  B_i(t', {\bf x}')\Big]| \Psi\rangle
     \langle p_{\rm out},\alpha_{\rm out}|\CF_i (t',\hat {\bf R}_\chi -{\bf x}')| p_{\rm in},\alpha_{\rm in}\rangle\\
    & + \langle {\rm out} | \hat B_i({\bf x}')\Big[ \hat A(t,{\bf x}),\CF_i (t',\hat {\bf R}_\chi -{\bf x}')\Big] |{\rm in}\rangle\bigg\}\,.
\end{align}
The second term vanishes under the assumptions that $\hat{A}$ is a target observable and $\CF_i$ is an external force, thus $[\hat{A},\CF_i]=0$. 
Furthermore, we can write the first term using the correlation function Eq.~\eqref{eqApp:linresp, Dcorr} and the matrix element over $\CF_i$ using its Fourier transform $\CF_i(\vb{q})$, resulting in 
\begin{align} \label{eq:kubo_1}
    \langle \Delta \hat A \rangle (t,\vb{x}) = \int \dd t'\, {\cal D}_{A B_i}(t-t',\vb{x};-{\bf q})\,e^{-i \Delta E t'}\langle \alpha_{\rm out}|\CF_i(-{\bf q})|\alpha_{\rm in}\rangle 
    = e^{-i \Delta E t} {\chi}_{A B_i}(\Delta E,\vb{x};-{\bf q})\,\langle \alpha_{\rm out}|\CF_i(-{\bf q})|\alpha_{\rm in}\rangle\,.
\end{align}
Importantly, by ${\cal D}_{A B_i}(\tau,\vb{x};-{\bf q})$ we mean using Eq.~\eqref{eqApp:linresp, Dcorr} with $\hat B=\hat B_i (-\vb{q})$, the spatial Fourier transform of $\hat B_i(\vb{x}')$. 
Finally, a Fourier transformation with respect to $\vb{x},\,t$ results in the Kubo formula~\cite{Kubo:1957mj}
\begin{align} \label{eqApp:linresp,DeltaAinChi}
    \langle \Delta \hat A \rangle(\omega, {\bf q};-\vb{q}) = (2\pi) \, \delta (\omega - \Delta E)\chi_{AB_i} (\omega,{\bf q}; -{\bf q})\,\langle \alpha_{\rm out}|\CF_i(-{\bf q})|\alpha_{\rm in}\rangle\,.\nonumber\\
\end{align}
Looking at Eq.~\eqref{eqApp:linresp,DeltaAinChi}, it is clear that $\chi_{AB_i} (\omega,{\bf q}; -{\bf q})$ characterizes the response of the target due to the interaction with the external force. Moreover, comparing Eq.~\eqref{eqApp:linresp,DeltaAinChi} with Eq.~\eqref{eqApp:linsresp,DAinChiAHint}, it is clear that
\begin{align} \label{eqApp:linresp, Chirels}
  \chi_{A \langle H_{\rm int}\rangle}(\omega, {\bf q}; -{\bf q}) = \chi_{AB_i} (\omega,{\bf q}; -{\bf q})\,\langle \alpha_{\rm out}|\CF_i(-{\bf q})|\alpha_{\rm in}\rangle\,.  
\end{align}

A key result is given by the fluctuation-dissipation theorem. 
Consider the target to be in thermal equilibrium, described by the density matrix $\hat \rho_{\rm \Psi}$. Given a target operator $\hat B$, the associated structure function is defined to be
\begin{align} \label{eqApp:linresp,Sdef}
    &S_{B}(\omega) = \int \dd \tau\,\Tr{\hat \rho_{\Psi} \hat{B}^{\dagger}(\tau) \hat B }\,e^{i\omega \tau} \ .
\end{align}
One can show that \cite{Boothroyd:2020}
\begin{align} \label{eqApp:linsresp, flucdiss}
    S_{B}(\omega)   = -2 [1+f_{\rm BE}(\omega)] \Im \chi_{B^\dagger B}(\omega)\,,
\end{align}
where $f_{\rm BE}(\omega)=1/(e^{\omega/T}-1)$ is the Bose-Einstein distribution. 
Note that the left-hand side of Eq.~\eqref{eqApp:linsresp, flucdiss} refers to the dynamic part of the correlation function and is strictly valid only  for $\omega \neq 0$, whereas the DC component can be accounted for separately with a $\delta(\omega)$ term.
Since we consider finite detection thresholds $\omega > 0$ we can safely omit this contribution to the material response.

A particularly useful case is when the operator $\hat B$ is chosen as $\hat B = \langle \hat H_{\rm int}\rangle$. 
Using the definition in Eq.~\eqref{eqApp:linresp,Sdef}, it can be shown~\cite{Boothroyd:2020} that $S_{\langle H_{\rm int}\rangle}\left(\Delta E,\vb{q}\right)$ corresponds to the Fermi Golden rule differential rate $\frac{\dd\Gamma}{\dd^3\vb{q}}$. 
In order to make energy conservation manifest, we work with the differential rate $\frac{\dd\Gamma}{\dd^3\vb{q}\dd\omega}$ --- obtained through multiplying by  $\delta (\omega-\Delta E)$ enforcing the kinematics --- which results in
\begin{align}
    \frac{\dd\Gamma}{\dd^3\vb{q}\dd\omega} = S_{\langle H_{\rm int}\rangle}\left(\omega,\vb{q}\right)\,(2\pi)\, \delta(\omega-\Delta E) = (1+f_{\rm BE}(\omega))\,(4\pi)\, \delta(\omega-\Delta E) \,\Im \left(-\chi_{\langle H_{\rm int}\rangle^\dagger \langle H_{\rm int}\rangle}(\omega,\vb{q};-\vb{q})\right)\,.
\end{align}
Since the interaction $\hat H_{\rm{int}}$ (Eq.~\eqref{eqApp:linresp,Hint}) is a convolution
of system and probe operators, the rate factorizes into target and probe contributions.
The susceptibility Eq.~\eqref{eqApp:linresp, Chirels} decomposes to
\begin{align}
    \chi_{\langle H_{\rm int}\rangle^\dagger \langle H_{\rm int}\rangle}(\omega,\vb{q};-\vb{q}) = \chi_{B_i^\dagger B_j} (\omega,\vb{q}; -{\bf q})\,\langle \alpha_{\rm out}|\CF_i(-{\bf q})|\alpha_{\rm in}\rangle^{*}\,\langle \alpha_{\rm out}|\CF_j(-{\bf q})|\alpha_{\rm in}\rangle\,, 
\end{align}
and similarly the differential rate is given by
\begin{align} \label{eqApp:linsresp,dGammadqdw}
    \frac{\dd\Gamma}{\dd^3\vb{q}\dd\omega} = (1+f_{\rm BE}(\omega))\, (4\pi) \,\delta(\omega-\Delta E) \,\Im \Big[-\chi_{B_i^\dagger B_j} (\omega, {\bf q}; -\vb{q})\,\langle \alpha_{\rm out}|\CF_i(-{\bf q})|\alpha_{\rm in}\rangle^{*}\,\langle \alpha_{\rm out}|\CF_j(-{\bf q})|\alpha_{\rm in}\rangle\Big]\,.
\end{align}

%%%%%%%%%%%%%%%%%%%%%%%%%%%%%%%%%%%%%%%%
\section{Derivation of the master formula}\label{app:master eq derivation}
%%%%%%%%%%%%%%%%%%%%%%%%%%%%%%%%%%%%%%%%
%
The master formula Eq.~\eqref{eq:master_eq} is a direct consequence of Eq.~\eqref{eqApp:linsresp,dGammadqdw} when the choice of $\{\hat B_i\}$ and $\{\CF_i\}$ is made according to Eq.~\eqref{eq:Hint}. Indeed, we can rewrite Eq.~\eqref{eq:Hint} in the suggestive form
\begin{align}
     \hat {\cal H}_{\rm int}(\vb{q}) = \hat{B}_\mu(\vb{q}) \CF_\mu(\vb{q})\,,
\end{align}
where $\hat{B}_0(\vb{q})=\hat n_e (\vb{q})$, $\hat{B}_i(\vb{q})=\hat S_e^{i} (\vb{q})$, $\CF_0(\vb{q})= \generalV \hat {F}_0(\vb{q})$, and $\CF_i(\vb{q})=\generalV \hat{F}_i(\vb{q})$. 
Throughout this appendix we use the indices $0 \leq \mu,\nu \leq 3$ and follow the summation convention.
We write them with lower indices to indicate that no raising, lowering, or contraction with the Minkowski metric is implied. 
Now, plugging this choice of operators along with the kinematic constraint $\Delta E = \omega_{\vb{q}}\equiv \vb{q} \cdot \vb{v}_{\chi} - \frac{q^2}{2m_\chi}$ into Eq.~\eqref{eqApp:linsresp,dGammadqdw} results in
\begin{align} \label{eqApp:linresp, maseqpolarized}
    \frac{\dd\Gamma_{\cal{O}}}{\dd^3\vb{q}\dd\omega} = |\generalV|^2 (1+f_{\rm BE}(\omega))\,(4\pi) \,  \delta(\omega-\omega_{\vb{q}}) \,\Im \Big[-\chi_{B_\mu^\dagger B_\nu} (\omega, {\bf q};-{\bf q})\,\langle \alpha_{\rm out}|\hat F_\mu(-{\bf q})|\alpha_{\rm in}\rangle^{*}\,\langle \alpha_{\rm out}|\hat F_\nu(-{\bf q})|\alpha_{\rm in}\rangle\Big]\,.
\end{align}
The subscript $\cal{O}$ signifies that the rate is associated with the effective operator $\cal{O}$, according to Eq.~\eqref{eq:form factors}. 
A particularly important case is when the DM spin is not polarized, corresponding to averaging Eq.~\eqref{eqApp:linresp, maseqpolarized} over the initial spin $\alpha_{\rm{in}}$ and summing over the final spin $\alpha_{\rm{out}}$, yielding
\begin{align}\label{eq:relativistic}
    \frac{\dd\Gamma_{\cal{O}}}{\dd^3\vb{q}\dd\omega} = |\generalV|^2 (1+f_{\rm BE}(\omega))\, (2\pi) \,\delta(\omega-\omega_{\vb{q}}) \,\Im \Big[-\chi_{B_\mu^\dagger B_\nu} (\omega, {\bf q}; -{\bf q})\,{\rm{Tr}}\Big\{{\hat F}_\mu \hat F_\nu^\dagger\Big\}\Big]\,.
\end{align}
We can also rewrite this expression in another form that separates the contributions of the force terms and the susceptibilities,
\begin{align} 
    \frac{\dd\Gamma_{\cal{O}}}{\dd^3\vb{q}\dd\omega} = |\generalV|^2 (1+f_{\rm BE}(\omega))\,(2\pi) \, \delta(\omega-\omega_{\vb{q}})\,\chi^{\prime\prime}_{B_\mu^\dagger B_\nu}(\omega, {\bf q};- {\bf q})\,{\rm{Tr}}\Big\{{\hat F}_\mu \hat F_\nu^\dagger\Big\}\,,
\end{align}
where
\begin{align}
\chi^{\prime \prime}_{B_\mu^\dagger B_\nu}(\omega, {\bf q}; -{\bf q})\equiv\frac{\chi_{B_\nu^\dagger B_\mu}^* (\omega, {\bf q}; -{\bf q}) - \chi_{B_\mu^\dagger B_\nu} (\omega, {\bf q}; -{\bf q})}{2i}
\end{align}
is the absorptive part of the susceptibility. The disadvantage of this expression is that it is not manifestly real, and so we choose to work with Eq.~\eqref{eq:relativistic}, while also separating the susceptibilities from the forces in a manifestly real way. 
To do this, we define the symmetric $(+)$ and antisymmetric $(-)$ susceptibilities 
\begin{equation}
    \chi^{\pm}_{B_\mu^\dagger B_\nu} (\omega, {\bf q}; -{\bf q})\equiv\frac{\chi_{B_\mu^\dagger B_\nu} (\omega, {\bf q}; -{\bf q})\pm\chi_{B_\nu^\dagger B_\mu} (\omega, {\bf q}; -{\bf q})}{2}\,,
\end{equation}
using which we get
\begin{align} \label{eqApp:linsresp, Mastereqcomp} 
    \frac{\dd\Gamma_{\cal{O}}}{\dd^3\vb{q}\dd\omega} =& |\generalV|^2 (1+f_{\rm BE}(\omega))\,(2\pi)\, \delta(\omega-\omega_{\vb{q}})
    \nonumber\\
    &\times \Bigg[{\rm{Im}}\left(-\chi^+_{B_\mu^\dagger B_\nu} (\omega, {\bf q}; -{\bf q})\right)\,{\rm{Re}}\left({\rm{Tr}}\Big\{\hat F_\mu \hat F_\nu^\dagger\Big\}\right)
    +{\rm{Re}}\left(-\chi^{-}_{B_\mu^\dagger B_\nu} (\omega, {\bf q}; -{\bf q})\right)\,{\rm{Im}}\left({\rm{Tr}}\Big\{\hat F_\mu \hat F_\nu^\dagger\Big\}\right)\Bigg]\,.
\end{align}
In order to cast the equation into the form of Eq.~\eqref{eq:master_eq}, we separate the $\mu,\nu=0$ indices from the others in Eq.~\eqref{eqApp:linsresp, Mastereqcomp},
\begin{align} \label{eqApp:linsresp, Mastereqexp} 
    \frac{\dd\Gamma_{\cal{O}}}{\dd^3\vb{q}\dd\omega} &= |\generalV|^2 (1+f_{\rm BE}(\omega))\,(2\pi) \, \delta(\omega-\omega_{\vb{q}})
    \Bigg[{\rm{Im}}\left(-\chi_{n_e n_e} (\omega, {\bf q}; -{\bf q})\right)\,{\rm{Tr}}\Big\{\hat \SIff \hat \SIff^\dagger\Big\}\nonumber\\
    &+{\rm{Im}}\left(-\chi^+_{S_e^{i} S_e^{j}} (\omega, {\bf q}; -{\bf q})\right)\,{\rm{Re}}\left({\rm{Tr}}\Big\{\hat F_i \hat F_j^\dagger\Big\}\right)+{\rm{Re}}\left(-\chi^{-}_{S_e^{i} S_e^{j}} (\omega, {\bf q}; -{\bf q})\right)\,{\rm{Im}}\left({\rm{Tr}}\Big\{\hat F_i \hat F_j^\dagger\Big\}\right)\nonumber\\
    &+2\,{\rm{Im}}\left(-\chi^+_{n_e S_e^{i}} (\omega, {\bf q}; -{\bf q})\right)\,{\rm{Re}}\left({\rm{Tr}}\Big\{\hat \SIff \hat F_i^\dagger\Big\}\right)+2\,{\rm{Re}}\left(-\chi^{-}_{n_e S_e^{i}} (\omega, {\bf q}; -{\bf q})\right)\,{\rm{Im}}\left({\rm{Tr}}\Big\{\hat \SIff \hat F_i^\dagger\Big\}\right)\Bigg]\,,
\end{align}
where we used the fact that ${\rm{Tr}}\Big\{\hat \SIff \hat \SIff^\dagger\Big\}$ is real, and that $n_e,\,S_e$ are Hermitian.

To lighten the notation, we now define $\chi_{00} (\omega, {\bf q})\equiv\chi_{n_e n_e} (\omega, {\bf q}; -{\bf q})$, $\chi^\pm_{ij} (\omega, {\bf q})\equiv\chi^\pm_{S_e^{i} S_e^{j}} (\omega, {\bf q}; -{\bf q})$, and $\chi^{\pm}_{0i} (\omega, {\bf q})\equiv\chi^{\pm}_{n_e S_e^{i}} (\omega, {\bf q}; -{\bf q})$, and drop the hats from operators. 
With these new definitions, we can now write our master equation
\begin{align} \label{eqApp:linsresp, Mastereqfin} 
    \frac{\dd\Gamma_{\cal{O}}}{\dd^3\vb{q}\dd\omega} &= |\generalV|^2 (1+f_{\rm BE}(\omega))\,(2\pi) \, \delta(\omega-\omega_{\vb{q}})
    \Bigg[{\rm{Im}}\left(-\chi_{00} (\omega, {\bf q})\right)\,{\rm{Tr}}\Big\{ \SIff\SIff^\dagger\Big\}\nonumber\\
    &+{\rm{Im}}\left(-\chi^+_{ij} (\omega, {\bf q})\right)\,{\rm{Re}}\left({\rm{Tr}}\Big\{F_i F_j^\dagger\Big\}\right)+{\rm{Re}}\left(-\chi^{-}_{ij} (\omega, {\bf q})\right)\,{\rm{Im}}\left({\rm{Tr}}\Big\{ F_i F_j^\dagger\Big\}\right)\nonumber\\
    &+2\,{\rm{Im}}\left(-\chi^+_{0i} (\omega, {\bf q})\right)\,{\rm{Re}}\left({\rm{Tr}}\Big\{\SIff F_i^\dagger\Big\}\right)+2\,{\rm{Re}}\left(-\chi^{-}_{0i} (\omega, {\bf q})\right)\,{\rm{Im}}\left({\rm{Tr}}\Big\{\SIff F_i^\dagger\Big\}\right)\Bigg]\, ,
\end{align}
or in terms of the compact notation of Eq.~\eqref{eq:relativistic} as
\begin{align}
    \frac{\dd\Gamma_{\cal{O}}}{\dd^3\vb{q}\dd\omega} = |\generalV|^2 (1+f_{\rm BE}(\omega))\,(2\pi) \, \delta(\omega-\omega_{\vb{q}}) \,\Im \Big[-\chi_{\mu\nu} (\omega, {\bf q})\,{\rm{Tr}}\Big\{F_\mu F_\nu^\dagger\Big\}\Big]\,,
\end{align}
where we denoted $\chi_{\mu\nu} (\omega, {\bf q})\equiv\chi_{B_\mu^\dagger B_\nu} (\omega, {\bf q}; -{\bf q})$. 
To clarify when each term contributes, it is useful to classify the operators in Table~\ref{tab:NR_operators} into four sets.
The first two contain operators for which $\vb{F}=0$:  
\begin{equation}
    A_1 = \{ {\cal O}_1, {\cal O}_2, {\cal O}_8 \}, 
    \qquad
    A_2 = \{ {\cal O}_5, {\cal O}_{11} \}.
\end{equation}
The latter two contain operators for which $F_0 = 0$:  
\begin{equation}
    A_3 = \{ {\cal O}_4, {\cal O}_6, {\cal O}_7, {\cal O}_{12}, {\cal O}_{15} \}, 
    \qquad
    A_4 = \{ {\cal O}_3, {\cal O}_9, {\cal O}_{10}, {\cal O}_{13}, {\cal O}_{14} \}.
\end{equation}
The term ${\rm Im}(-\chi_{00})$ contibutes when considering operators from $A_1$ or $A_2$ and ${\rm Im}(-\chi_{ij}^+)$  contributes when considering operators from $A_3$ or $A_4$.
The terms ${\rm Re}(-\chi^-_{ij})$, ${\rm Im}(-\chi^+_{0i})$,  and ${\rm Re}(-\chi^-_{0i})$
contribute only when sums of operators from distinct sets are present: 
\begin{equation}
    \begin{array}{lcl}
    {\rm Re}(-\chi^-_{ij}) &:& A_3 \text{ with } A_4,\\[3pt]
    {\rm Im}(-\chi^+_{0i}) &:& A_1 \text{ with } A_3, \text{ or } A_2 \text{ with } A_4,\\[3pt]
    {\rm Re}(-\chi^-_{0i}) &:& A_1 \text{ with } A_4,  \text{ or } A_2 \text{ with } A_3.
    \end{array}
\end{equation}

%%%%%%%%%%%%%%%%%%%%%%%%%%%%%%%%%%%%%%%%%%%%%%%%%%%%%
\section{Mapping to Dielectric Function Formalism}\label{app:mapping to dielectric}
%%%%%%%%%%%%%%%%%%%%%%%%%%%%%%%%%%%%%%%%%%%%%%%%%%%%%
%
Here we review how the well-established dielectric function formalism for computing DM spin-independent  rates~\cite{Hochberg:2021pkt,Knapen:2021run,Boyd:2022tcn} appears within our framework.
Consider a probe of charge $Qe$ ({\it e.g.} free electron, DM) interacting with a target characterized by the electron number density $n_e(\vb{x})$.
In the non-relativistic limit, the spin-independent probe-electron scattering is described by the ${\cal O}_1$ operator, which mediates scalar interactions and constitutes the leading-order term (in a velocity expansion) for vector-mediated interactions.
Assuming a Coulomb interaction, we consider the following interaction Hamiltonian in the frequency space
\begin{align} \label{eqApp:Hinteqdiel}
    {\cal H}^{\rm int} (\vb{q}) = \frac{4\pi Q \alpha_{\rm em}}{q^2} n_e(\vb{q})\,,
\end{align}
corresponding to choosing $\SIff=1$, $\SDff=0$, and $V(q) = 4\pi Q \alpha_{\rm em} /q^2$ in Eq.~\eqref{eq:Hint}. Plugging this in our master formula Eq.~\eqref{eqApp:linsresp, Mastereqfin} yields the rate
\begin{align}
    \frac{\dd\Gamma_{\cal{O}}}{\dd^3\vb{q}\dd\omega} &= |V(q)|^2 (1+f_{\rm BE}(\omega))\,(4\pi) \, \delta(\omega-\omega_{\vb{q}})\,
    {\rm{Im}}\left(-\chi_{00} (\omega, {\bf q})\right)\,,
\end{align}
suggesting that $  {\rm{Im}}\left(-\chi_{0 0} (\omega, {\bf q})\right)$ should be related to the loss function. This is indeed the case, as we show below (see also Ref.~\cite{Nozieres:1959,Boyd:2022tcn}).

We compare Maxwell's macroscopic equations with the microscopic ones while making use of Eq.~\eqref{eqApp:linresp,DeltaAinChi} with a proper choice of the target observable $\hat A$. On the one hand, Maxwell's macroscopic equations in a dielectric medium read
\begin{align}
   {\boldsymbol \nabla} \cdot  {\bf D}  = Q\, e \, n_{\rm{probe}}\,,\quad\quad 
   {\boldsymbol \nabla} \times {\bf E} = -\frac{\partial {\bf B}}{\partial t}\,,
\end{align}
where $n_{\rm{probe}}$ and $Q e$ are the number density and charge of the probe respectively.
The contribution of the magnetic field in the non-relativistic limit is ${\cal{O}}(v)$ and so we neglect it. Taking the Fourier transform of the equations results in
\begin{align}
   i {\bf q} \cdot {\boldsymbol{\epsilon}}\cdot {\bf E} =  Q \,e \, n_{\rm{probe}}(\omega,{\bf q})\,,\quad\quad 
   i {\bf q} \times {\bf E} = {\mathcal{O}(v)} \approx 0 \,,
\end{align}
where we used the fact that the displacement field ${\bf{D}}$ is related to ${\bf E}$ and the dielectric tensor $\boldsymbol{\epsilon}$ as
\begin{align} \label{eqApp: macmaxeq}
     {\bf D}(\omega, {\bf q}) = \boldsymbol\epsilon (\omega, {\bf q})\cdot {\bf E}(\omega, {\bf q})\,.
\end{align}
Neglecting the velocity-suppressed transverse component of ${\bf E}$ we find
\begin{align} \label{eqApp: macmaxeq2}
   i q \,\epsilon_L  E_L =  Q\, e \, n_{\rm{probe}}(\omega,{\bf q})\,,
\end{align}
where $\epsilon_L\equiv \hat{\bf{q}}\cdot\boldsymbol\epsilon \cdot\hat{\bf{q}}$ is the longitudinal dielectric function and $E_L = \vu{q} \cdot \vb{E}$ is the the logitudinal electric field.
On the other hand, the microscopic Maxwell’s equation for $\vb{E}$ states that it is sourced by both the density of the probe $n_{\rm{probe}}$ and the induced fluctuations of the electron density $\Delta n_e$ in the target,
\begin{align} \label{eqApp: micmaxeq}
    i {\bf q} \cdot {\bf E} = iq\,E_L = Q\,  e\,  n_{\rm{probe}}(\omega,{\bf q}) + e \, \Delta n_e(\omega,{\bf q)}\,.
\end{align}
Combining Eqs.~\eqref{eqApp: macmaxeq2} and~\eqref{eqApp: micmaxeq} results in
\begin{align} \label{eqApp:dielecrel1}
    \frac{1}{\epsilon_L(\omega, {\bf q})} = 1 +\frac{\Delta n_e(\omega,{\bf q)}}{Q \, n_{\rm{probe}}(\omega,{\bf q})}\,.
\end{align}
We now use Eq.~\eqref{eqApp:linresp,DeltaAinChi} with the choice $\hat A = \hat n_e$ and $\alpha_{\rm in}=\alpha_{\rm out}$, specialized to our choice of the interaction Hamiltonian in Eq.~\eqref{eqApp:Hinteqdiel}, resulting in
\begin{align} \label{eqApp:linresp,Deltan}
    \langle \Delta n_e \rangle(\omega, {\bf q}) = (2\pi)\, \delta (\omega - \omega_{\bf{q}})\,\frac{4\pi Q \alpha_{\rm em}}{q^2}\,\chi_{0 0} (\omega,{\bf q})\,,
\end{align}
Identifying $n_{\rm{probe}}(\omega,{\bf q}) = (2\pi)\, \delta (\omega - \omega_{\bf{q}})$ and plugging Eq.~\eqref{eqApp:linresp,Deltan} in Eq.~\eqref{eqApp:dielecrel1}, we obtain
\begin{align} \label{eqApp:dielecrel2}
    \frac{1}{\epsilon_L(\omega, {\bf q})} = 1 +\frac{4\pi \alpha_{\rm em}}{q^2}\,\chi_{00} (\omega,{\bf q})\,.
\end{align}
Finally, taking the imaginary part establishes the relation between the loss function and $  {\rm{Im}}\left(\chi_{00} (\omega, {\bf q})\right)$,
\begin{align} \label{eqApp:dielecrel2}
    {\rm{Im}}\left(-\frac{1}{\epsilon_L(\omega, {\bf q})}\right) = \frac{4\pi \alpha_{\rm em}}{q^2}\,{\rm{Im}}\left(-\chi_{00} (\omega, {\bf q})\right)\,.
\end{align}

%%%%%%%%%%%%%%%%%%%%%%%%%%%%%%%%%%%%%%
\section{Praseodymium Spin Density Response}\label{app:Praseodymium}
%%%%%%%%%%%%%%%%%%%%%%%%%%%%%%%%%%%%%%
%
\begin{figure}
    \centering
    \includegraphics[width=0.47\linewidth]{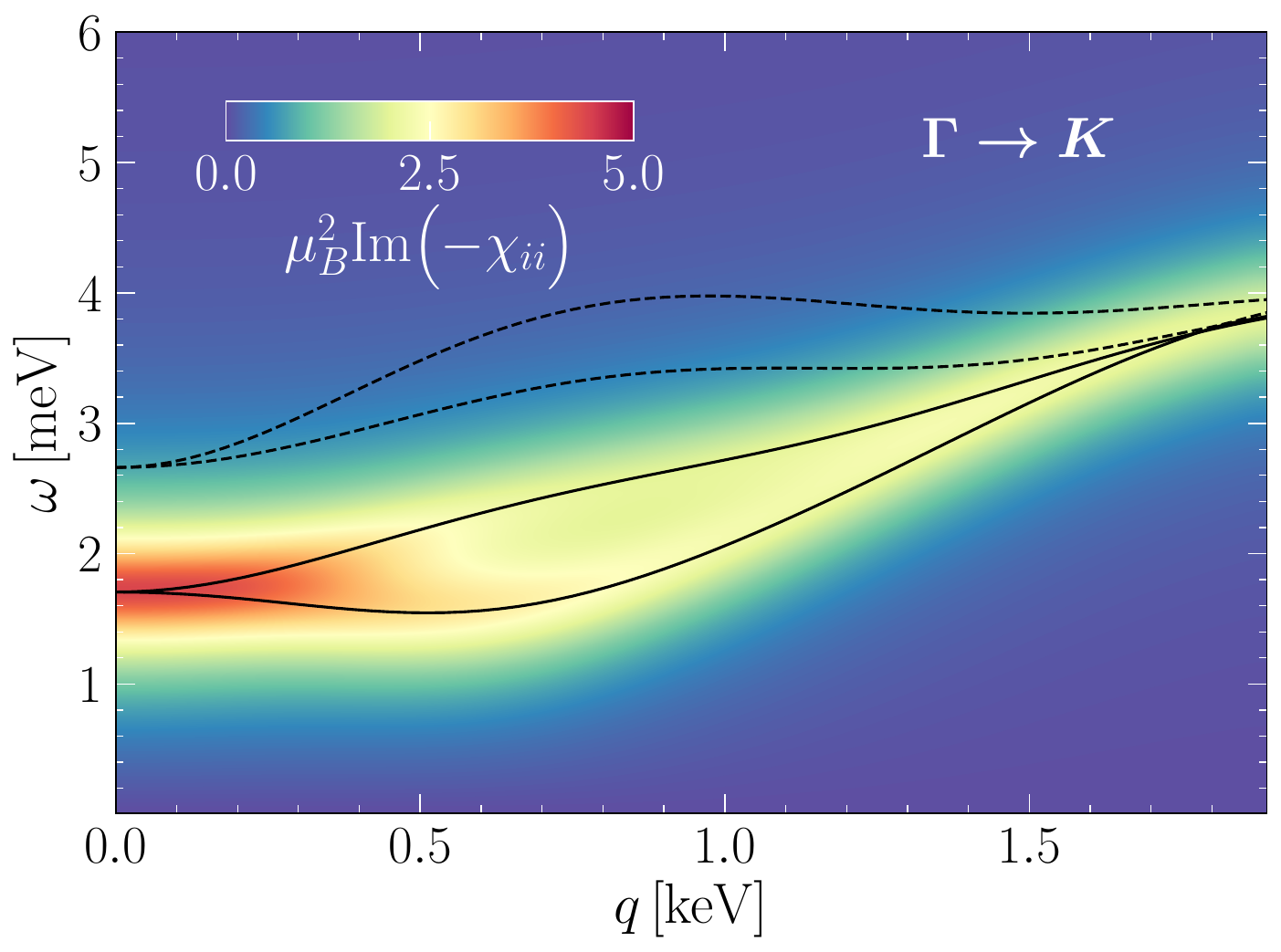}  
\hspace{0.5cm}
\includegraphics[width=0.47\linewidth]{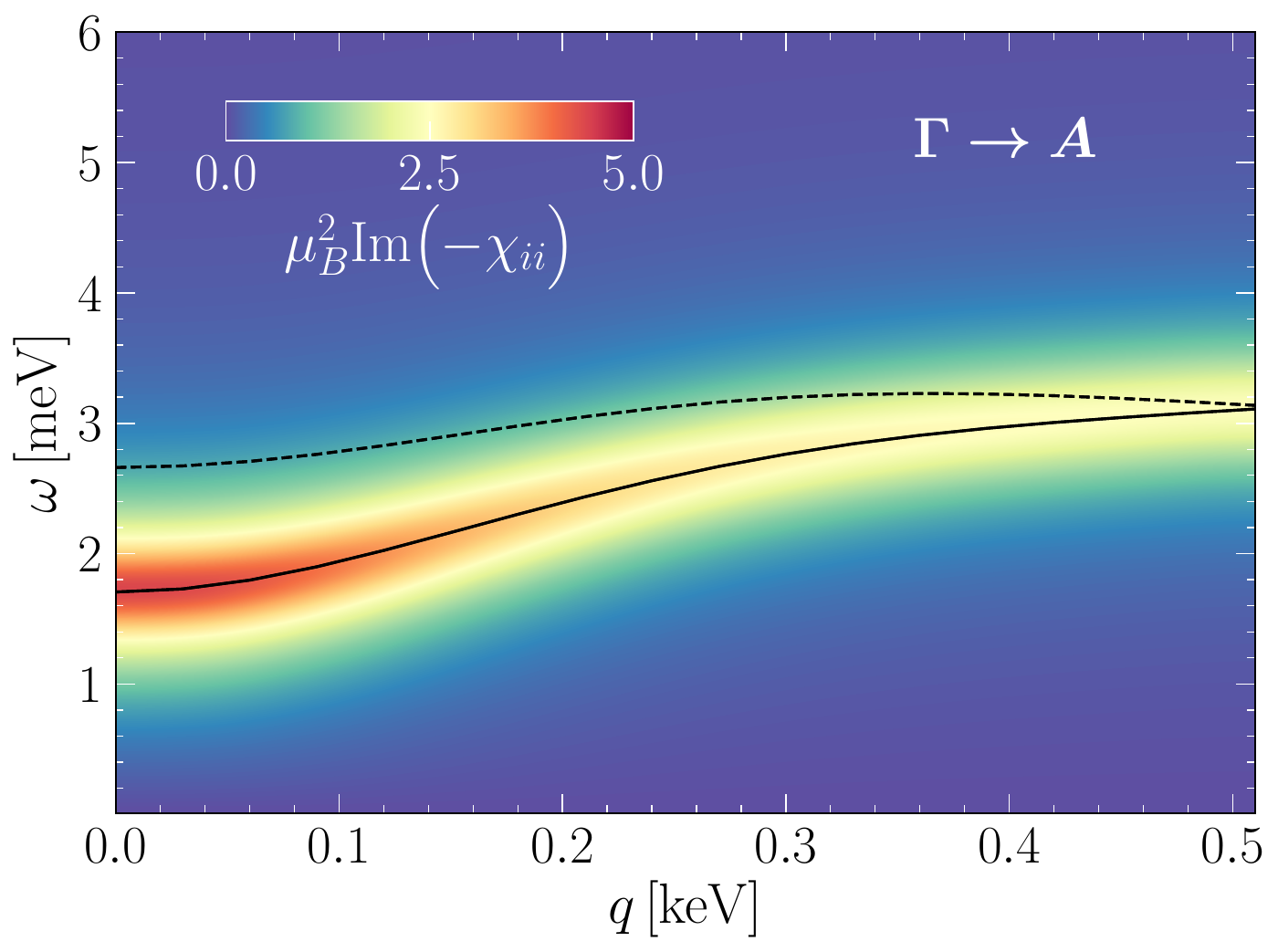}
\includegraphics[width=0.47\linewidth]{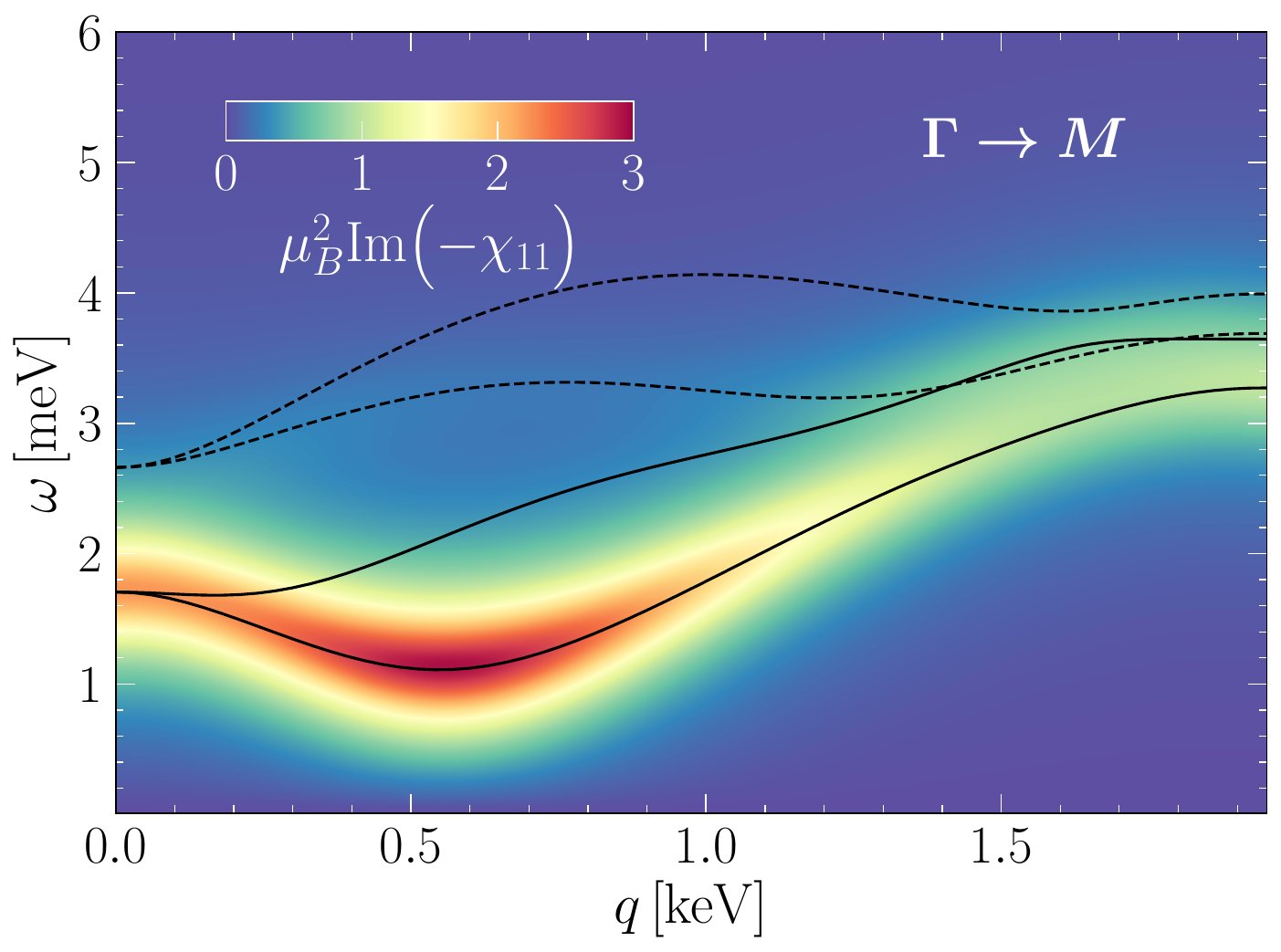}  
\hspace{0.5cm}
\includegraphics[width=0.47\linewidth]{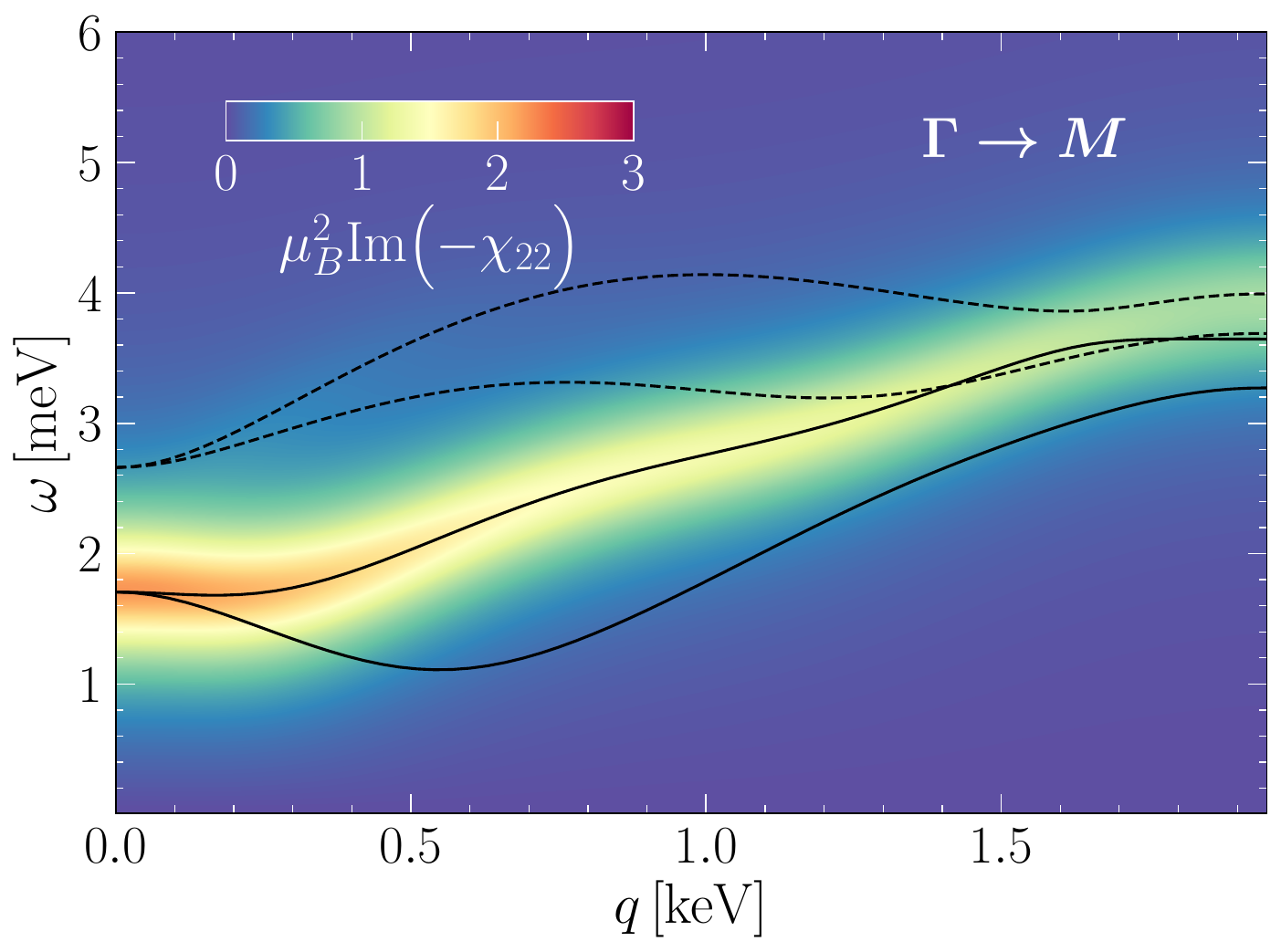}
    \caption{The spin response function for Pr computed using the MF-RPA susceptibility Eq.~\eqref{eq:Pr response} 
    with the two-ion couplings extracted from a fit of the dispersion relations Eq.~\eqref{eq:Pr dispersion  relation} in neutron scattering experiments \cite{Bak:1975,Houmann:1975,Houmann:1979} with a width $\Gamma=0.45$~meV.
    The upper left and upper right panels show the trace of the spin response, ${\rm Im}(\chi_{ii})$, for momentum deposits along the $\Gamma \to K$ and $\Gamma \to M$ directions, respectively.
    The lower left and lower right panels present the individual components, ${\rm Im}(-\chi_{11})$ and ${\rm Im}(-\chi_{22})$, for momentum deposits along the $\Gamma \to M$ direction.
    The difference between the upper panels highlights the dependence of the spin susceptibility $\chi_{ij}$ on momentum direction, while the variation between the lower panels illustrates differences between its components.
    Black curves indicate the dispersion relations of the different modes.  
    }
    \label{fig:pr_response_appendix}
\end{figure}
In this appendix, we describe the analytical modeling we use for the spin response of the rare earth metal Praseodymium (Pr), discussed in Section~\ref{ssec:spin} and for which we have computed the projections shown in Fig.~\ref{fig:spin response bounds}.
We consider a commercially available crystal of Pr atoms arranged in a double hexagonal close-packed lattice.
The spin response of this crystal has been measured in neutron scattering experiments~\cite{Houmann:1971, Bak:1975, Houmann:1979}, finding remarkable agreement with a simple modeling of the spin dynamics of the outer shell $4f^2$ electrons in the ground state.
For a comprehensive review, see Ref.~\cite{Boothroyd:2020}.

Two thirds of the Pr ions lie in two equivalent hexagonal sub-lattices denoted by the indices $a,b$ (this factor of $2/3$ rescales the material density $\rho_{\rm T}$ in Eq.~\eqref{eq:R formula}).
In the absence of spin-interactions, the ground state and the next doublet of excited states are separated by a gap $\Delta = 3.5\, {\rm meV}$.
The electron spin interactions are described by the Heisenberg Hamiltonian
\begin{eqnarray}
    H = - \frac{1}{2} \sum_{\alpha \beta} J_{\alpha \beta}^{ij} S^i_{\alpha}  S^j_{\beta},
\end{eqnarray}
where $J_{\alpha \beta}^{ij}$ are couplings of order $10\, \mu {\rm eV}$ found by matching to experimental data~\cite{Houmann:1971, Bak:1975, Houmann:1979}.
Using the mean-field (MF) approximation and random-phase-approximation (RPA) one finds that the spin couplings break the degeneracy of the doublet into four energy bands
\begin{align}\label{eq:Pr dispersion  relation}
    \Omega^{\pm}_i({\bf q}) = \sqrt{\Delta ^2 - j(j+1)\Delta [J_{11}^{ii}({\bf q}) \pm |J_{12}^{ii}({\bf q}) |]}\ \, , \qquad i = {x,y}\, ,
\end{align}
where $J^{ij}_{ab}(\vb{q}) = V \, \sum_{\alpha\in a} \sum_{\beta \in b} J^{ij}_{\alpha \beta} e^{i \vb{q} \cdot (\vb{r}_\alpha - \vb{r}_\beta)}$ is the discrete Fourier transform of the couplings along the sub-lattices, $j=4$ is the total angular momentum of the states and $V = (2\rho_{\rm Pr} / 3m_{\rm Pr})^{-1}$ is the volume of a hexagonal unit cell.
The hexagonal lattice structure leads to different distances between spins in the $\vu{x}$ and $\vu{y}$ directions and thus different couplings. 
Similarly, the sub-lattice structure in the $\vu{z}$ directions causes couplings in the same sub-lattice $J_{11}(\vb{q})$ and in different sub-lattices $J_{12}(\vb{q})$ to differ.
Overall, the doublet is split into four bands differing by a few meV.

The electronic spin-susceptibility can be analytically calculated in the MF-RPA approximation and is given by
\begin{eqnarray}\label{eq:Pr response}
    \chi_{lm}(\omega, \vb{q}) & = & \frac{2\, \chi_{lm}^0(\omega, \vb
    {q}) \bqty{1 - \chi_{ll}^0(\omega, \vb{q}) (J^{ll}_{11}(\vb{q}) - {\rm Re}\, J^{ll}_{12}(\vb{q}))}}
    {1 - \chi_{ll}^0(\omega, \vb{q}) \pqty{J^{ll}_{11}(\vb{q}) - \abs{J^{ll}_{12}(\vb{q})}}} \, , \\
    \chi_{lm}^0(\omega, \vb{q}) & \equiv &  \frac{1}{V} \frac{g^2\, j(j+1) \, \Delta \, \abs{f(q)}^2}{\omega^2 - \Delta^2 + i \omega \Gamma} (\delta_{lm} - \delta_{l3} \delta_{m3}) \, ,
\end{eqnarray}
where we do not use the index summation convention in Eq.~\eqref{eq:Pr response}. 
Here $\chi_{lm}^0$ is the spin-susceptibility of each sub-lattice in the absence of spin couplings $J^{ij}_{\alpha\beta}$, $g=4/5$ is the Landé $g$-factor, $f(q)$ is the measured magnetic form factor~\cite{Lebech1979} of Pr, and $\chi_{lm}$ is the MF-RPA spin-susceptibility accounting for the spin couplings.
Neutron scattering spectra show the width $\Gamma$ varies between $0.45\, {\rm meV}$ at $q=400 \, {\rm eV}$ to $\sim 2 \, {\rm meV}$ at $q=0$.
We also note that a convenient approximation for the susceptibility is given by the $\Gamma \to 0$ limit
\begin{eqnarray}
    \chi_{ij}(\omega,\vb{q}) =\pqty{\delta_{ij} -\delta_{i3}\delta_{j3}} \, \sum_{s \in \Bqty{\pm}} \frac{i \pi}{2} \frac{g^2 \, j(j+1) \Delta \, \abs{f(q)}^2}{V \, \Omega^s_i(\vb{q})} \pqty{ 1 + \frac{{\rm Re} \, J_{12}(\vb{q})}{\abs{J_{12}(\vb{q})}}} \, \delta\pqty{\omega - \Omega^{s}_i(\vb{q})} \, .
\end{eqnarray}

In Figs.~\ref{fig:pr_response} and \ref{fig:pr_response_appendix}, we present the trace of the Pr crystal spin response, ${\rm Im}(-\chi_{ii})$, with $\chi_{ij}$ modeled according to Eq.~\eqref{eq:Pr response}. 
The response is evaluated along several crystal directions, with solid black lines indicating the underlying energy bands. 
It is peaked at small momenta and energies $\sim 2\, {\rm meV}$, in addition to exhibiting anisotropic dependence on the momentum direction $\vu{q}$.

%%%%%%%%%%%%%%%%%%%%%%%%%%%%%%%%%%%%%%%
\section{Auxiliary Calculations --- Phase Space Integrals}\label{app:auxiliary calculations}
%%%%%%%%%%%%%%%%%%%%%%%%%%%%%%%%%%%%%%%%
%
In all the bounds and projections shown in the manuscript, we assume the DM velocity distribution follows the Standard Halo Model with velocity dispersion $v_0 = 220\, {\rm km}/{\rm s}$, earth velocity in the galactic frame $v_{\oplus} = 232 \, {\rm km}/{\rm s}$ and escape velocity $v_{\rm esc} = 540 \, {\rm km}/{\rm s}$~\cite{Lewin:1995rx,Stanic:2025yze}.
Explicitly, the velocity distribution is a Boltzmann distribution truncated at $v_{\rm esc}$,
\begin{eqnarray}
    f_{\rm h} (\vb{v}_{\rm h}) &= N_0^{-1} e^{-v_{\rm h}^2/v_0^2} \Theta(v_{\rm esc}^2 - v_{\rm h}^2) \, , \qquad
    N_0 = \pi^{3/2}
    v_0^3 \bigg[ {\rm erf}\left(\frac{v_{\rm esc}}{ v_0}\right) - \frac{2}{\sqrt{\pi} } \frac{v_{\rm esc}}{v_0} e^{-v_{\rm esc}^2/v_0^2 }\bigg]\ ,
\end{eqnarray}
where $v_{\rm h}$ is the DM velocity in the galactic frame.
A simple Galilean transform converts the halo velocity $\vb{v}_{\rm h}$ distribution to the lab velocity distribution $\vb{v}$ (the frame where the target is at rest): $f_{\rm h}({\bf v} + {\bf v}_{\oplus}(t))$, {\it i.e.} shifting the halo velocity by the Earth velocity in the galactic frame. 
Note that the direction of $\vb{v}_\oplus$ modulates daily according to the rotation of the earth around its axis. 
The velocity of the DM wind in the lab frame is given by $\vb{v} = \vb{v}_{\rm h} - \vb{v}_{\oplus}(t)$.

Calculating the DM-electron interaction rates requires multi-dimensional integration over velocities, energies, momenta, and angles. 
In this appendix, we provide auxiliary calculations for various moments of the velocity distribution that can be analytically calculated in advance, thereby reducing runtime.
The full kinematic distribution is given by
\begin{eqnarray}
    f(\vb{v}, \omega, \vb{q}) =  f_{\rm h}(\vb{v} + \vb{v}_{\oplus}) \, \delta\pqty{\omega - \pqty{\vb{q} \cdot \vb{v} -  \frac{q^2}{2m_\chi}}} \,  \Theta(\omega)
    \newrow{,}
\end{eqnarray}
where $\Theta$ is the Heaviside theta function.
In all moments, we average over $\omega$ and use the notation $\ev{\cdot }_X$ to delineate averaging over parameters $X$.

We start with moments of the transverse velocity $\vb{v}_{\perp} \equiv \pqty{1 - \vu{q} \vu{q}^T} \cdot \vb{v} = {\bf v}- ({\bf v}\cdot \vu{q}) \vu{q}$
\begin{eqnarray}
    g_n^{i_1, i_2, ..., i_n}(\omega,{\bf q},t) \equiv \ev{v^{i_1}_{\perp} \cdots v^{i_n}_{\perp}}_{\vb{v}} =
    \int \dd^3 {\bf v}_\chi f(\vb{v}, \omega, \vb{q}) v^{i_1}_{\perp} \cdots v^{i_n}_{\perp}  \, .
\end{eqnarray}
The simplest moment is
\begin{eqnarray}
    g_0(\omega,{\bf q},t) & = &\frac{\pi v_0^2}{q N_0}
    \bqty{e^{-x_-^2}  -e^{-x_{\rm esc}^2}} \Theta(\omega) \Theta(v_{\rm esc} - v_-) \, ,
\end{eqnarray}
where
\begin{eqnarray}
    v_{-} & \equiv & v_{\rm min} + \vu{q} \cdot \vb{v}_{\oplus} \, , \qquad  v_{\rm min} = \frac{\omega}{q} + \frac{q}{2 m_\chi} \, , \qquad x \equiv \frac{v}{v_0} \, .
\end{eqnarray}
The velocity $v_{\rm min}$ is the minimal velocity required for scattering with energy transfer $\omega$ and momentum transfer $q$. 
The kinematics ensure $v_- < v_{\rm esc}$, which enforces integration limits on the momentum angle $\mu \equiv \vu{q} \cdot {\bf\hat v}_{\oplus}$ 
\begin{eqnarray}
    \mu_{\rm max}
    = \min \left\{ 1, \frac{1}{v_{\oplus}} \left[  v_{\rm esc} - v_{\rm min} \right] \right\} \, , \qquad
    \mu_{\rm min}
    = \max \left\{ -1, -\frac{1}{v_{\oplus}} \left[  v_{\rm esc} + v_{\rm min}\right] \right\} \, . 
\end{eqnarray}
By direct calculation, we find that
\begin{align}
    \vb{g}_1(\omega, \vb{q}, t) = -\vb{v}_{\oplus} \,  g_0 + \vu{q} \pqty{\mu v_{\oplus} \,  g_0  - \ev{\vb{v}_{\rm h} \cdot \vu{q}}_{\vb{v}_{\chi}}} \, ,
\end{align}
and
\begin{eqnarray}\label{eq:g2}
    g_2^{ij}(\omega,\vb{q},t) & = &  \pqty{{v}_\oplus^i {v}_\oplus^j - \mu v_{\oplus} \, \pqty{v_{\oplus}^i   {\hat q}^j + {v_{\oplus}}^j   {\hat q}^i} }g_0 + \frac{1}{3} \delta^{ij} \ev{\vb{v}_{\rm h} \cdot \vu{q}}_{\vb{v}}
    + \bqty{
    \ev{\pqty{\vb{v}_{\rm h} \cdot \vu{q}}^2}_{\vb{v}} 
    + \mu^2 v_{\oplus}^2 \, g_0 - \frac{2}{3}  \ev{v_{\rm h}^2}_{\vb{v}} }
    {\hat q}^i  {\hat q}^j
    \newrow{,}
\end{eqnarray}
where
\begin{eqnarray}
    \ev{\vb{v}_{\rm h} \cdot \vu{q}}_{\vb{v}} & = & \frac{\pi v_0^3 x_-}{N_0 q}\bigg[ e^{-\frac{v_-^2}{v_0^2}}-e^{-\frac{v_{\rm esc}^2}{v_0^2}}\bigg] \Theta(\omega) \Theta(v_{\rm esc} - v_-) \, ,
    \newrow{}
    \ev{\pqty{\vb{v}_{\rm h} \cdot \vu{q}}^2}_{\vb{v}} & = & \frac{\pi v_0^4 x_-^2}{N_0 q}
    \bigg[ e^{-x_-^2}-e^{-x_{\rm esc}^2}\bigg] \Theta(\omega) \Theta(v_{\rm esc} - v_-) \, , \\[5pt]
     \ev{v_{\rm h}^2}_{\vb{v}} & = & \frac{\pi v_0^4}{N_0 q}\bqty{e^{-x_-^2} \left(x_-^2 + 1 \right) - e^{-x_{\rm esc}^2} \left(x_{\rm esc}^2 + 1\right) } \Theta(\omega) \Theta(v_{\rm esc} - v_-) \, . \nn 
\end{eqnarray}

We proceed with $\vu{q}$ angle-averaged moments, which can be used assuming the material responses are isotropic.
We align $\vu{z} \parallel \vb{v}_{\oplus}$ and introduce the abbreviated notation
\begin{eqnarray}
    \eta_{n m}^{i_1\ldots i_n j_1 \ldots j_m}(\omega, q) = \ev{v^{i_1}_{\perp} \cdots v^{i_n}_{\perp} \hat{q}^{j_1} \cdots \hat{q}^{j_m}}_{\vb{v}, \, \Omega_{\vb{q}}} \, .
\end{eqnarray}
For $n=0$ the non-vanishing components up to $m=4$ are given by
\begin{eqnarray}
    \eta_{00} & = & \ev{1}_{\vb{v}, \, \Omega_{\vb{q}}}\,,
\end{eqnarray}
\begin{eqnarray}
    \eta^{xx}_{02} & \equiv & \eta^{yy}_{02} = \frac{1}{2} \ev{1}_{\vb{v}, \, \Omega_{\vb{q}}} - \ev{\mu^2}_{\vb{v}, \, \Omega_{\vb{q}}} \, , \qquad \eta^{zz}_{02} = \ev{\mu^2}_{\vb{v}, \, \Omega_{\vb{q}}} \, ,
\end{eqnarray}
\begin{eqnarray}
    \eta_{04}^{xxyy} & = & \eta_{04}^{xyxy} = \eta_{04}^{xyyx} = \eta_{04}^{yyxx} = \frac{1}{3}\eta_{04}^{xxxx} = \frac{1}{3}\eta_{04}^{yyyy} 
    = \frac{1}{8} \pqty{\frac{1}{2} \ev{1}_{\vb{v}, \, \Omega_{\vb{q}}} - 2\ev{\mu^2}_{\vb{v}, \, \Omega_{\vb{q}}} + \ev{\mu^4}_{\vb{v}, \, \Omega_{\vb{q}}} } \, 
    \newrow{,}
    \eta_{04}^{xxzz} & = & \eta_{04}^{xzxz} = \eta_{04}^{yyzz} = \eta_{04}^{yzyz} = \eta_{04}^{zyzy} = \eta_{04}^{zzxx} = \eta_{04}^{zzyy} = \frac{1}{2} \pqty{
    \ev{\mu^2}_{\vb{v}, \, \Omega_{\vb{q}}} - \ev{\mu^4}_{\vb{v}, \, \Omega_{\vb{q}}} } 
    \, ,\\[5pt]
    \eta_{04}^{zzzz} & = & \ev{\mu^4}_{\vb{v}, \, \Omega_{\vb{q}}}\, , \nn 
\end{eqnarray}
where
\begin{eqnarray}
    \ev{1}_{\vb{v}, \, \Omega_{\vb{q}}} & = & \frac{\pi^2 v_0^2}{N_0 q x_{\oplus}} 
    \bqty{\sqrt{\pi} {\rm Erf}(x_-) 
    - 2 e^{-x_{\rm esc}^2} x_{\oplus} \mu }_{\mu_{\rm min}}^{\mu_{\rm max}} 
    \newrow{}
    \ev{\mu^2}_{\vb{v}, \, \Omega_{\vb{q}}}  & = &  \frac{\pi^2 v_0^2}{2 N_0 q x_{\oplus}^3} 
    \bqty{\sqrt{\pi}\pqty{1 + 2 x_{\rm min}^2} {\rm Erf}(x_-) 
    + 2 \pqty{2 x_{\rm min} - x_-} e^{-x_-^2} 
    - \frac{4}{3} e^{-x_{\rm esc}^2} x_{\oplus}^3 \mu^3}_{\mu_{\rm min}}^{\mu_{\rm max}}
    \, ,\\[5pt]
    \ev{\mu^4}_{\vb{v}, \, \Omega_{\vb{q}}} & = &  \frac{\pi^2 v_0^2}{4 N_0 q x_{\oplus}^5} \bigg[ 
    \sqrt{\pi}\pqty{3 + 12 x_{\rm min}^2 + 4 x_{\rm min}^4} {\rm Erf}(x_-) 
    - \frac{8}{5} e^{-x_{\rm esc}^2} x_{\oplus}^5 \mu^5
   \newrow{}
    && \qquad\qquad\qquad + 2 \pqty{8 x_-^2 x_{\min }-12 x_- x_{\min }^2+8 x_{\min }^3+8 x_{\min }-2 x_-^3-3
   x_-} e^{-x_-^2} 
    \bigg]_{\mu_{\rm min}}^{\mu_{\rm max}} \, . \nn
\end{eqnarray}

An additional moment that was used is
\begin{eqnarray}
    \eta_{20}^{xx} & = & \eta_{20}^{yy} = \frac{1}{2} \ev{(\vb{v}_{\rm h} \cdot \vu{q})^2}_{\vb{v}, \, \Omega_{\vb{q}}} - \frac{1}{2} \ev{(\vb{v}_{\rm h} \cdot \vu{q})^2 \, \mu^2}_{\vb{v}, \, \Omega_{\vb{q}}} + \frac{1}{3} \ev{v_{\rm h}^2 \mu^2}_{\vb{v}, \, \Omega_{\vb{q}}} + \frac{1}{2} v_\oplus^2 \pqty{\ev{\mu^2}_{\vb{v}, \, \Omega_{\vb{q}}} - \ev{\mu^4}_{\vb{v}, \, \Omega_{\vb{q}}}}
    \newrow{,}
    \eta_{20}^{zz} & = & \ev{(\vb{v}_{\rm h} \cdot \vu{q})^2 \, \mu^2}_{\vb{v}, \, \Omega_{\vb{q}}} + \frac{1}{3} \ev{v_{\rm h}^2}_{\vb{v}, \, \Omega_{\vb{q}}} - \frac{2}{3} \ev{v_{\rm h}^2 \mu^2}_{\vb{v}, \, \Omega_{\vb{q}}} + v_\oplus^2 \pqty{2 \ev{1}_{\vb{v}, \, \Omega_{\vb{q}}} - 2 \ev{\mu^2}_{\vb{v}, \, \Omega_{\vb{q}}} + \ev{\mu^4}_{\vb{v}, \, \Omega_{\vb{q}}}} \, ,
\end{eqnarray}
where
\begin{eqnarray}
    \ev{v_{\rm h}^2}_{\vb{v}, \, \Omega_{\vb{q}}} & = & \frac{\pi^2 v_0^4}{2 N_0 q x_\oplus} \bqty{ 3 \sqrt{\pi} {\rm Erf}(x_-) - 2 e^{-x_-^2} x_- - 4 x_\oplus \mu (1 + x_{\rm esc}^2) e^{-x_{\rm esc}^2}  }_{\mu_{\rm min}}^{\mu_{\rm max}}
    \newrow{,}
    \ev{v_{\rm h}^2 \, \mu^2}_{\vb{v}, \, \Omega_{\vb{q}}} & = & \frac{\pi^2 v_0^4}{12 N_0 q x_\oplus^3} 
    \bqty{ 3 \sqrt{\pi} (5 + 6 x_{\rm min}^2) {\rm Erf}(x_-) + 6 (3 x_- - 2 x_- x_\oplus^2 \mu^2 - 8 x_\oplus \mu)  e^{-x_-^2} 
    - 8 x_\oplus^3 \mu^3 (1 + x_{\rm esc}^2) e^{-x_{\rm esc}^2}  }_{\mu_{\rm min}}^{\mu_{\rm max}} 
    \newrow{,}
    \ev{(\vb{v}_{\rm h} \cdot \vu{q})^2}_{\vb{v}, \, \Omega_{\vb{q}}} & = & \frac{\pi^2 v_0^4}{6 N_0 q x_\oplus} 
    \bqty{ 3 \sqrt{\pi} {\rm Erf}(x_-) 
    - 6 x_-  e^{-x_-^2} 
    - 4 x_\oplus (3 x_{\rm min}^2 \mu + 3 x_\oplus x_{\rm min} \mu^2 + x_\oplus^2 \mu^3 ) e^{-x_{\rm esc}^2}  }_{\mu_{\rm min}}^{\mu_{\rm max}} 
    \newrow{,}
    \ev{(\vb{v}_{\rm h} \cdot \vu{q})^2 \mu^2}_{\vb{v}, \, \Omega_{\vb{q}}} & = & \frac{\pi^2 v_0^4}{12 N_0 q x_\oplus^3} 
    \bigg[ 3 \sqrt{\pi} (3 + 2 x_{\rm min}^2) {\rm Erf}(x_-) 
    +6 ( 4 x_\oplus \mu + x_-(2 x_\oplus^2 \mu^2 - 1) )  e^{-x_-^2} 
    \newrow{}
    && \qquad\qquad\qquad\qquad\qquad\qquad
    - \frac{4}{5} x^3_\oplus (10 x_{\rm min}^2 \mu^3 + 15 x_\oplus x_{\rm min} \mu^4 + 6 x_\oplus^2 \mu^5 ) e^{-x_{\rm esc}^2}  \bigg]_{\mu_{\rm min}}^{\mu_{\rm max}} \, . 
\end{eqnarray}

\bibliography{references.bib}

\end{document}